\documentclass{article}

\usepackage{arxiv}

\usepackage[utf8]{inputenc} 
\usepackage[T1]{fontenc}    
\usepackage{hyperref}       
\usepackage{url}            
\usepackage{booktabs}       
\usepackage{amsfonts}       
\usepackage{nicefrac}       
\usepackage{microtype}      
\usepackage{lipsum}		
\usepackage{graphicx}
\usepackage{natbib}
\usepackage{doi}

\usepackage{graphicx}
\usepackage{fancyhdr}
\usepackage{times}
\usepackage{bm}
\usepackage{bbm}
\usepackage{amsmath, amssymb, enumitem}
\usepackage{tikz}
\usetikzlibrary{automata, positioning, arrows, shapes.geometric}
\usepackage{caption}
\usepackage{subcaption}
\usepackage{multirow}
\usepackage{lscape} 
 \usepackage{rotating}

\title{Multistate models as a framework for estimand specification in clinical trials of complex processes}

\author{ Alexandra Bühler \\
    Statistics and Actuarial Science \\
	University of Waterloo \\
	Waterloo, Canada \\
	\texttt{abuhler@uwaterloo.ca} \\
	\And
	Richard J. Cook \\
	Statistics and Actuarial Science \\
	University of Waterloo \\
	Waterloo, Canada \\
	\texttt{rjcook@uwaterloo.ca} \\
	\And 
	Jerald F. Lawless \\
	Statistics and Actuarial Science \\
	University of Waterloo \\
	Waterloo, Canada \\
	\texttt{jlawless@uwaterloo.ca} \\
}

\date{}
\begin{document}
\maketitle

\begin{abstract}
Intensity-based multistate models provide a useful framework for characterizing disease processes, the introduction of interventions, 
loss to followup, and other complications arising in the conduct of randomized trials studying complex life history processes. Within this framework we discuss the issues involved in the specification of estimands and show the limiting values of common estimators of marginal process features based on cumulative incidence function regression models.
When intercurrent events arise we stress the need to carefully define the target estimand and the importance of avoiding targets of inference
that are not interpretable in the real world. This has implications for analyses, but also the design of clinical trials where protocols may help in the interpretation of estimands based on marginal features.
\end{abstract}


\section{Introduction}
\subsection{Overview}
The randomized clinical trial is widely regarded as the best study design for evaluating an experimental intervention compared to standard care 
on disease processes.
However when the processes involve multiple types of events, recurrent events, or competing terminal events, it is often unclear how best to summarize treatment effects. In particular it is challenging to specify a sufficiently informative one-dimensional estimand to summarize effects and be used as a basis for tests. In cancer trials, for example, patients are at risk of cancer progression and death, and there has been considerable discussion about the merits of responses based on progression alone, progression-free survival and overall survival \citep{booth2012}. 

In cardiovascular trials, interest lies in prolonging overall survival but with recent advances in medical therapies and surgical interventions mortality rates are now relatively low. This has lead to increased interest in using responses based on recurrent myocardial infarction, stroke, and hospitalization, in addition to time to cardiovascular-related death; non-cardiovascular deaths are typically handled as competing risks \citep{Rufibach2019,Furberg2021, Toenges2021}. The conceptualization and observation of treatment effects can be further complicated by intercurrent events \citep{qu2021}, defined as events that  prevent observation of the primary response or otherwise interfere with the process of interest.
Early withdrawal from a clinical trial is an example of the former, while the introduction of rescue therapy is an example of the latter. Ways to define estimands and make treatment comparisons in the presence of intercurrent events have received a great deal of attention in recent years \citep{Rufibach2019}. Numerous researchers and working groups have proposed estimands and guidelines for specific settings involving competing risks \citep{Rufibach2019,  Stensrud2020, Young2020, Nevo2020, Poythress2020}, 
recurrent and terminal events \citep{Andersen2019, Schmidli2021, wei2021}, 
introduction of rescue therapies \citep{ster2020} or treatment switching \citep{watkins2013,Manitz2022}; see also \cite{casey2021} and \cite{stensrud2021}.

Discussion in the estimands literature regarding challenges in conceptualizing and specifying treatment effects  often lacks detail concerning models and assumptions,  which makes interpretation of the recommended estimands difficult. The challenges arise from the complexity of  disease processes and related processes involved with the management of subjects in a randomized clinical trial, and from the desire to make causal statements concerning treatment effects that are based on comparisons of marginal process features, which are not dependent on post-randomization events. In addition, estimands that reflect causal effects without a meaningful interpretation in the real world in which the trial is conducted are frequently considered. The use of potential outcomes has proven powerful and popular for formalizing causal reasoning and analysis, but can lead to causal statements about effects in randomized trials that do not reflect the actual trial or patient care.

Our goals are two-fold. First, we argue for the importance of models that reflect the disease process and related events that impact the conduct of the trial in both the planning and analysis stages.  Time is a basic element in trials on disease processes and so models based on stochastic processes are crucial; we discuss the particular utility of multistate models \citep{Andersen1993, Cook2018}. Our second goal is to present guiding principles for the specification of estimands for clinical trials involving complex disease processes. We consider settings involving competing risks and semi-competing risks in some detail, along with more complex processes involving multiple and possibly recurrent events. For the planning stage of a trial when primary and secondary analysis strategies are considered, we emphasize the important role of models that account for the 
disease process, study protocol, and clinical interventions.
In settings involving  intercurrent events, we show how multistate models 
can be used  to jointly represent disease and intercurrent event processes.
The structure offered by this multistate representation can 
clarify the interpretation of potential target estimands, 
reveal what is needed to estimate them, and illuminate whether they have an interpretation in the real world. We also highlight the value of specifying utilities for different disease states. While some may view these as subjective and undesirable for assessing the effects of experimental treatments in clinical trials, they are used to great effect in 
quality of life and health economic analyses \citep{Gelber1989, Torrance1986}. Moreover, utilities are often adopted implicitly; for example, 
progression-free survival and other composite time-to-event responses equally weight the component outcomes, and relative utilities of different disease paths are implied when computing  win ratios \citep{oakes2016}. 
Explicit specification and incorporation of utilities into the evaluation of treatments both enables the handling of complex  processes and makes explicit the relative importance of different events. 

The paper is organized as follows. In Section \ref{sec1.2} we describe challenges arising in two illustrative multicenter clinical trials.  
In Section \ref{sec2.0} we introduce notation for multistate models using illness-death and competing risks processes as illustrations,
define intensity functions, and give examples of functionals of process intensities which may serve as a basis for defining estimands. 
Principles for the specification of estimands are given in Section \ref{sec2.1}. In Section \ref{sec3} we discuss these processes in more detail, and provide some numerical results concerning the effects of assumptions, 
model misspecification and the interpretation of estimands. Section \ref{sec4} illustrates the utility of multistate models for dealing with more complex processes including those with intercurrent events.  Section \ref{sec5} contains remarks on potential outcomes and on utility-based estimands, and 
concluding remarks are given in Section \ref{sec6}.

\subsection{Some motivating and illustrative studies} \label{sec1.2} 
\subsubsection{Skeletal complications in patients with cancer metastatic to bone} \label{bonemets-study}

\begin{figure}[!ht] \begin{center}
\includegraphics[scale=0.65]{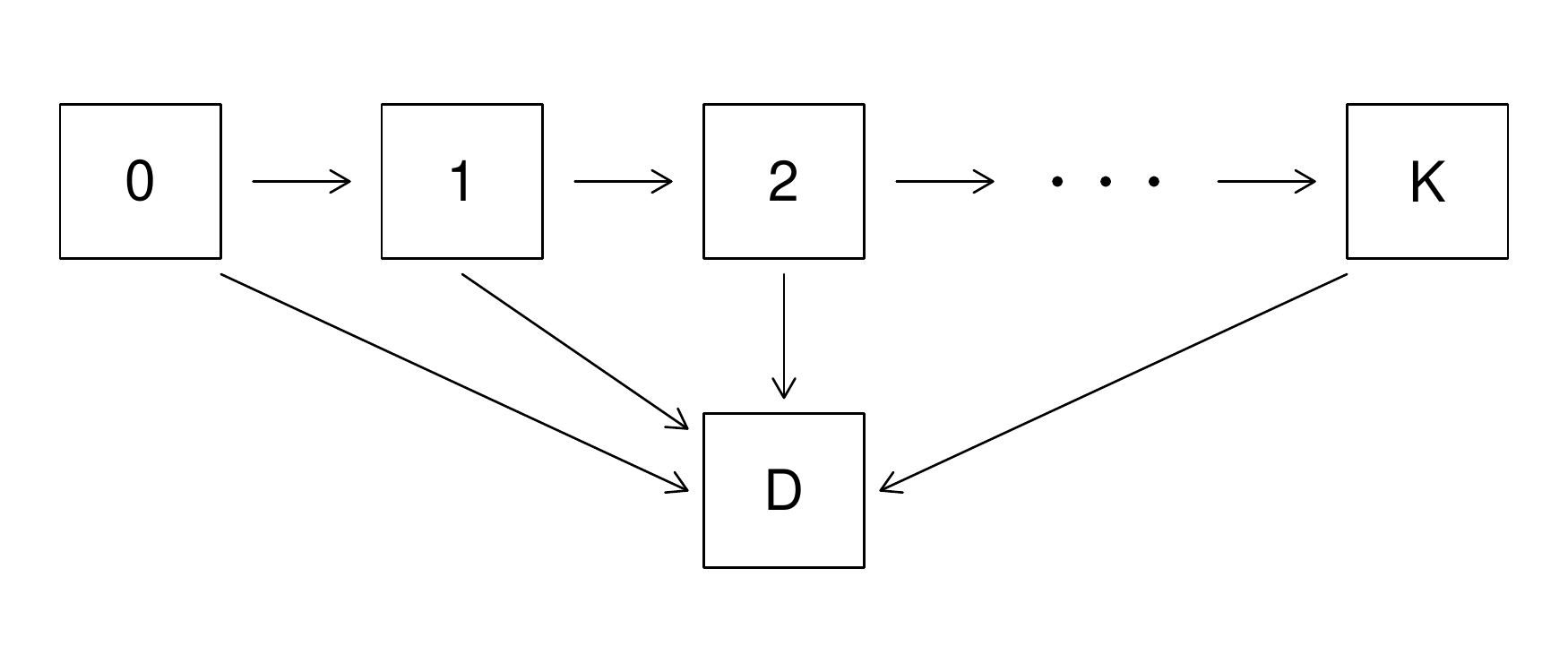}
\caption{A multistate model for up to $K$ recurrent skeletal complications and death for \cite{Theriault1999}.} \label{sre}
\end{center} \end{figure}

Many individuals with cancer experience skeletal metastases which can in turn cause fractures, bone pain, and the need for therapeutic or surgical 
interventions \citep{coleman2006}. Prophylactic treatment with a bisphosphonate can strengthen bone and reduce the risk of skeletal-related events and so are often used to mitigate complications and improve quality of life in affected individuals. \cite{Theriault1999} report on a international multicenter randomized trial involving patients with stage IV breast cancer having at least two predominantly lytic bone lesions greater than one centimeter in diameter. Eighty-five sites in the United States, Canada, Australia and New Zealand recruited patients who were randomized within strata defined by ECOG status to receive a 90 mg infusion of pamidronate every four weeks ($n=182$) or a placebo control ($n=189$). Skeletal complications of primary interest included nonvertebral and vertebral fractures,  the need for surgery to treat or prevent fractures, and 
the need for radiation for the treatment of bone pain.  
Individuals  with metastatic cancer are at high risk of death and here the skeletal-related event process is terminated by death, 
creating a semi-competing risks problem for the analysis of the time to the first skeletal event.
If interest lies in assessing the effect of pamidronate on the occurrence of recurrent skeletal complications then the 
recurrent event process is terminated by death.
Figure \ref{sre} is a multistate diagram depicting the possible occurrence of up to $K$ skeletal-related events while accommodating death from any of the transient
states.
Intensity-based analyses are natural for modeling such disease processes but marginal rate-based methods are more suited for primary analysis in clinical trials; we address this in detail in subsequent sections.
\cite{Cook2009} discuss methods for nonparametric estimation of rate and mean functions in this setting, while accommodating the possible impact of 
dependent censoring; we discuss approaches to analysis of the recurrent and terminal event processes in Section \ref{sec4.3} but discuss issues for the 
simpler illness-death process in Section \ref{sec3}.

\subsubsection{Assessing carotid endarterectomy versus medical care in stroke prevention} \label{nascet-intro}

Atherosclerosis involves the development of arterial plaque and stenosis of the carotid arteries putting individuals at increased risk of stroke.  \cite{Barnett1998} report on a multicenter clinical trial designed to evaluate the effect of a surgical intervention called carotid endarterectomy,
which involves surgical removal of plaque to increase the diameter of the carotid artery and enhance blood flow, compared to usual medical therapy 
(e.g.  use of platelet inhibitors and antihypertensive drugs).
The goal of both treatment strategies is the prevention of stroke and stroke-related death. To be eligible for recruitment individuals must have experienced at least one transient ischemic attack or minor stroke and have at least a 30\% narrowing of the carotid artery on the same side as (ipsilateral to) the event, as determined by central angiographic examination. Consenting patients were randomized to either carotid endarterectomy and best medical care, or best medical care alone; as part of best medical care, aspirin was recommended for all patients at a dose of 1,300 mg/day and blood pressure was carefully controlled through regular monitoring. The endpoints of interest include
i) any stroke arising from the same side as the one designated at the time of recruitment as a possible site for surgery,
ii) any stroke, 
iii) the composite event of stroke or stroke-related death, and
iv) the composite event of any stroke or death.
For  iii),  deaths unrelated to stroke are a competing risk which should be addressed when assessing the effect of surgery.

\begin{figure}[!ht]
\begin{center}
\includegraphics[scale=0.50]{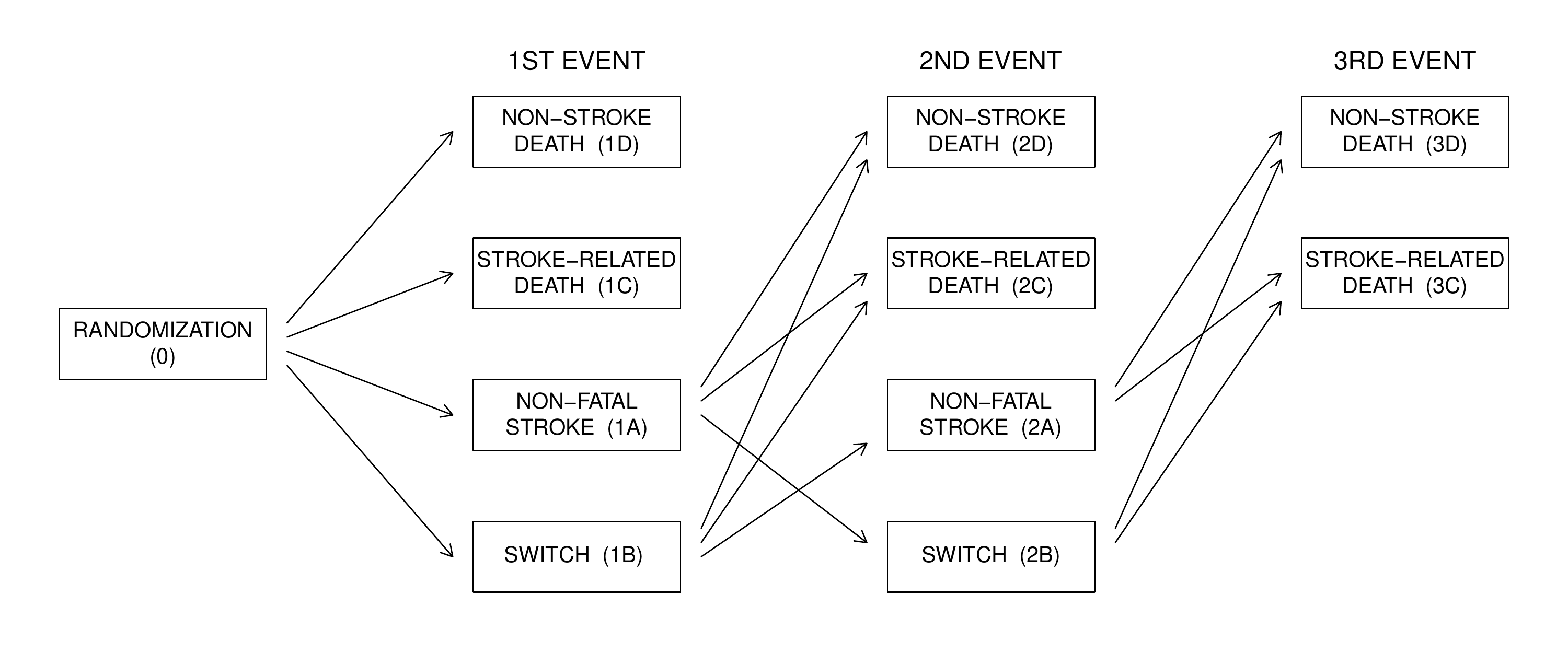}
\caption{A multistate model depicting the occurrence of stroke, switch from medical care to carotid endarterectomy, stroke-related death and non-stroke related death
for the stroke prevention trial reported on by \cite{Barnett1998}.}
\label{nascet}
\end{center}
\end{figure}
 
There were two strata of particular interest defined by whether the degree of carotid stenosis at randomization was moderate (stenosis of 30-69\%) or 
severe (stenosis of 70-99\%). An interim analysis led to early termination of recruitment to the severe stenosis stratum and publication of the finding that carotid endarterectomy was highly effective in the prevention of stroke among these patients \citep{NASCET-NEJM1991}. The recruitment and follow-up of the moderate stenosis stratum continued, but many individuals randomized to medical care ultimately underwent carotid endarterectomy,
further complicating analysis and interpretation of this data.
There were various reasons reported for crossover from the medical to surgical interventions including atherosclerotic progression from
moderate stenosis at randomization to severe steonsis during follow-up, at which point the criteria for carotid endarterectomy in the publication based on the severe 
stenosis stratum were met.
Physicians treating patients experiencing non-fatal stroke may also recommend that they undergo carotid endarterectomy to reduce the risk of future stroke.
In what follows we focus on the occurrence of the first stroke of any type and death due to stroke, and consider strategies for dealing with complications 
due to both the competing risk of death unrelated to stroke and participants randomized to medical care undergoing carotid endarterectomy.
Figure \ref{nascet} contains a multistate diagram depicting the occurrence of the first event among stroke, stroke-related death, death unrelated to stroke, 
and crossover to surgery; the latter event can only occur among patients randomized to medical care. 
Additional states can be entered upon the occurrence of subsequent events; see Figure \ref{nascet}.
We revisit this example in Section \ref{secnascet}. 

\section{Estimands for interventions in event history processes} \label{sec2}
\subsection{Multistate models, intensities and associated functionals} \label{sec2.0}

In a typical phase III  clinical trial, treatment groups are formed by randomizing individuals to receive either an experimental or control treatment. Although the disease process may be complex, a primary objective for estimation and testing of treatment efficacy is to compare some process feature in the two treatment groups. A process feature is typically related to some outcome or event, for example, the total time spent in a given state or the time of entry to a specific state. In that case a feature is based on the response distribution, for example,  the mean or median time to an event or the probability it occurs by a specific time. 
We define an estimand as a one-dimensional measure of the difference in a process feature in the experimental and control groups. 
The central clinical challenge involves specification of the process feature of greatest relevance, and then a clinical and  statistical challenge is to specify how to specify an estimand comparing the feature in the two groups.

\begin{figure}[!ht]
\begin{center}
\noindent
\begin{minipage}[t]{0.49\textwidth}
\begin{center}
\includegraphics[scale=0.50]{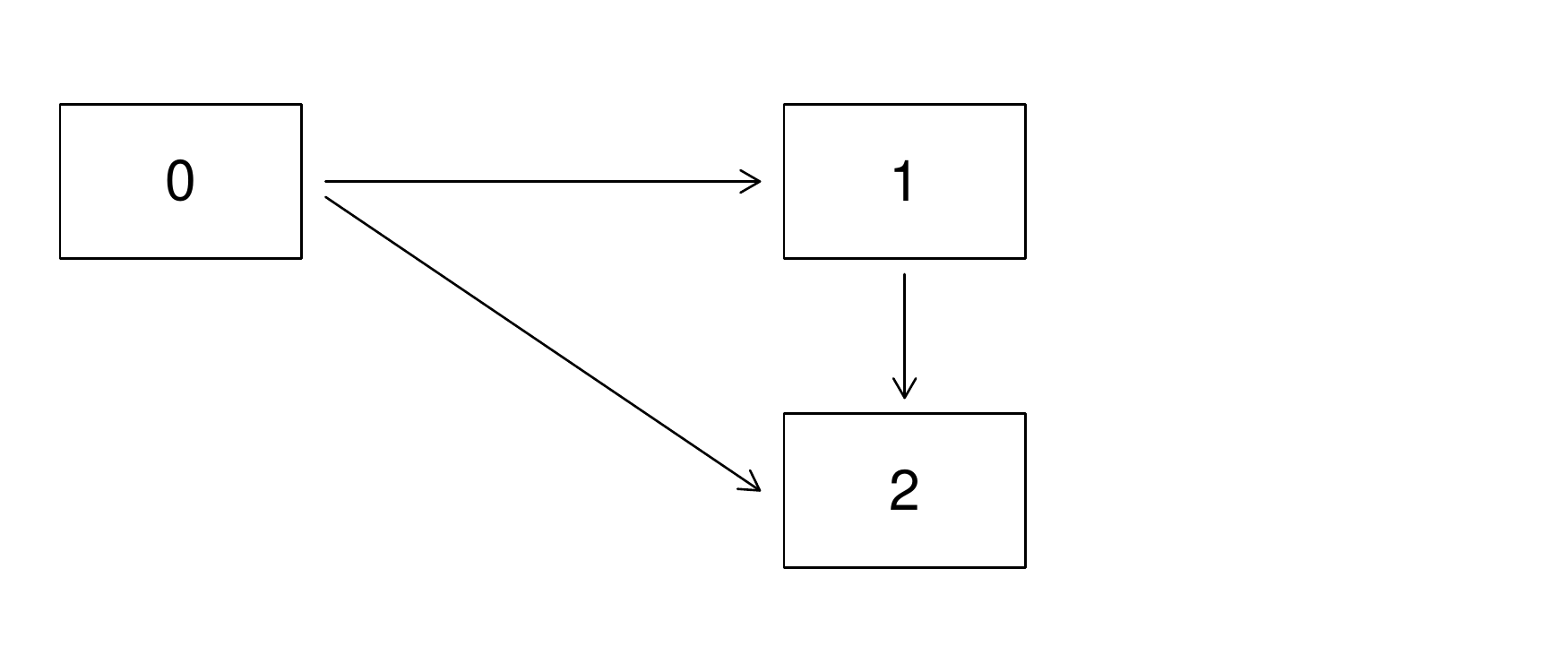} \\
{(a) An illness-death process}
\end{center}
\end{minipage}
\hfill
\noindent
\begin{minipage}[t]{0.49\textwidth}
\begin{center}
\includegraphics[scale=0.50]{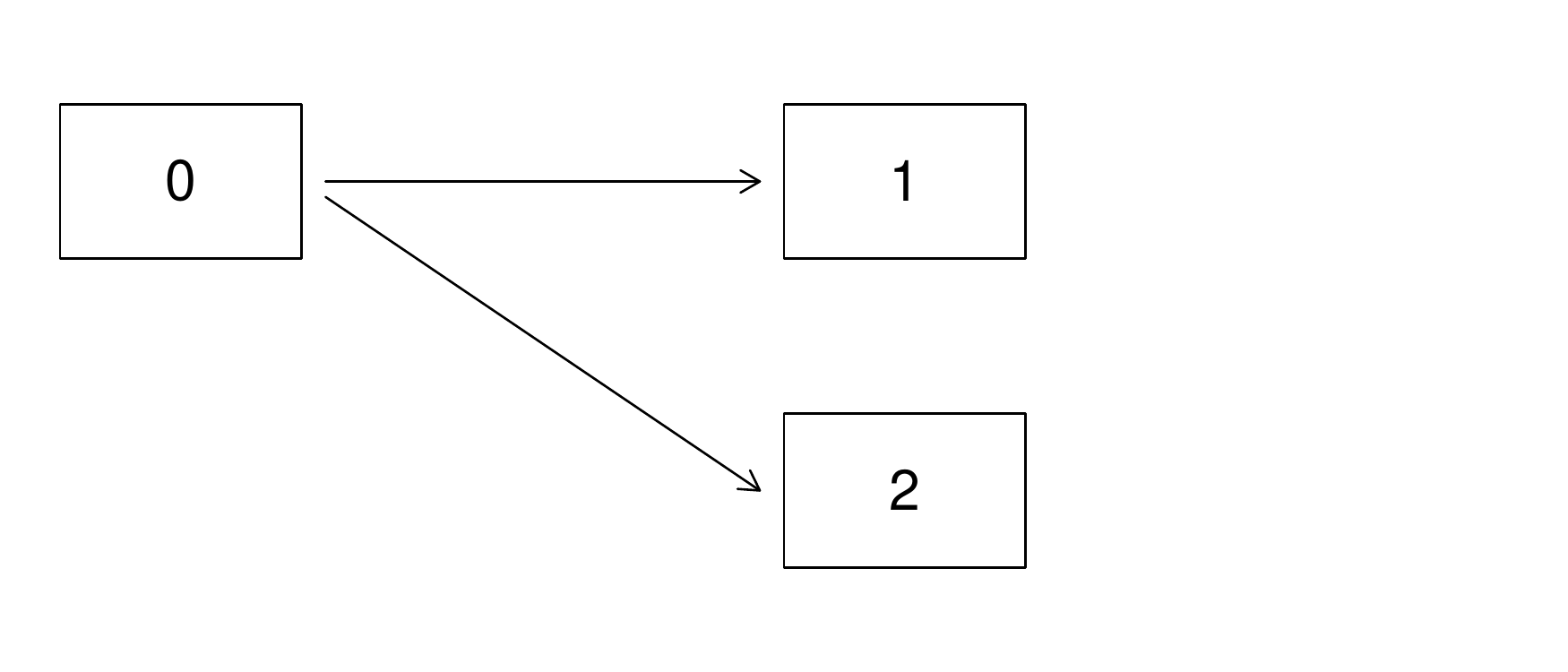} \\
 {(b) A competing risks process}
\end{center}
\end{minipage}
  \caption{Multistate diagrams for illness-death and competing risks processes.}
  \label{multistateDiagram}
\end{center}
\end{figure}

We consider disease processes that can be represented by multistate models and will 
first define notation, using the important illness-death process in Figure \ref{multistateDiagram}(a) with state space $\mathcal{S} = \{ 0, 1, 2 \}$ for illustration \citep{Andersen1993, Cook2018}. The related competing risks process shown in Figure \ref{multistateDiagram}(b) is discussed later.  The illness-death process is widely applicable to oncology trials (e.g. \citealp{carey2021}), with state 0 representing a cancer-free state following initial treatment, state 1 representing events such as recurrence,  relapse or progression, and state 2 representing death; individuals may enter state $2$ with or without having experienced the non-fatal event. 
We let $Z(t)$ denote the state occupied by a generic individual at time $t \geq 0$ and assume that the process $\{Z(t), t \ge 0\}$ begins in state $0$ at $t=0$,
the time of randomized treatment assignment.
We let $Y_{k}(t) = I(Z(t^-)=k)$ be a state occupancy indicator that  equals $1$ if an individual is in state $k \in \mathcal{S}$ at time $t^-$ and $0$ otherwise, and let  $T := \inf \{t>0: Z(t) \neq 0 \}$, $T_1 :=\inf \{t>0: Z(t)=1 \}$ and $T_2 := \inf \{t>0: Z(t) = 2 \}$ represent the event-free survival time (or time spent in state $0$), the time to the non-fatal event and the overall survival time (or time to death), respectively.  

Let $X$ be a  binary variable indicating whether an individual was randomized to receive the experimental ($X=1$) or control ($X=0$) treatment and let $V$ denote 
other observable covariates that may affect the disease process. 
For simplicity we initially assume that $V$ is fixed, but we later consider  settings where it may have time-varying components. If $\mathcal{H}(t)=\{Z(u): 0 \leq u < t, X,V \}$ is the history of the  process up to time $t^{-}$ plus  the treatment indicator $X$ and covariates $V$, the transition intensity function for a $k \longrightarrow l$ transition is defined as 
\begin{align}
\lim_{\Delta t \downarrow 0}  \dfrac{P(Z(t+\Delta t^{-})=l \ | \ Z(t^{-})=k, \mathcal{H}(t))}{\Delta t} = \lambda_{kl}(t \ | \ \mathcal{H}(t)),  \ \ \ \ \ \ \ \ k,l \in \mathcal{S}, \ \ k<l \; .
\end{align}
The stochastic nature of the multistate process is fully defined by the specification of all transition intensities.

For ease of discussion we consider Markov processes, in which case the intensities depend only on $t$ and the state occupied at $t^{-}$: 
$\lambda_{kl}(t \ | \ \mathcal{H}(t)) = Y_k(t)\lambda_{kl}(t \ | \ X,  V)$. 
We use the term \textit{process feature} to mean any functional of the set of intensities \citep{Andersen2012.1}. 
The survivor function for the event-free survival time is, for example, given by 
\begin{equation}
\label{EFS}
P(T > t|X,V) = P_{00}(0,t|X,V) = \exp \biggl( - \int_{0}^{t} (\lambda_{01}(u|X,V) + \lambda_{02}(u|X,V))du \biggr), 
\end{equation}
where  $P_{kl}(s,t \ | \ X,V) = P(Z(t)=l \ | \ Z(s)=k, X,V)$ for $s \le t$.
In addition $P(T_2 > t|X,V) = P_{00}(0,t|X,V) + P_{01}(0,t|X,V)$ where $P_{01}(0,t|X,V)$ is 
\begin{align*}
    \int_{0}^{t} P_{00}(0,u|X, V) \; \lambda_{01}(u|X,V)  \exp \biggl( - \int_{u}^{t} \lambda_{12}(s|X,V)ds \biggr) du \; .
\end{align*}

The distribution function for $T_1$ is often called the cumulative incidence function for the non-fatal event and is given by
\begin{equation} 
\label{CIF} 
 F_{1}(t|X,V) = P(T_1 \leq t|X,V) = \int_{0}^{t} P_{00}(0,u|X,V) \lambda_{01}(u|X,V)du \; .
\end{equation}
It should be noted that $F_1(t|X,V)$ approaches a limit less than one as $t$ increases.

\subsection{Principles for defining estimands} \label{sec2.1}

{
\cite{Andersen2012.1} provide important guidance on the analysis of life history processes to ensure interpretable results.
This has shaped our views on the specification of estimands and is recommended reading for those working in life history analysis in observational or 
experimental settings.
The three main tenets of \cite{Andersen2012.1} are: 
\textit{don't condition on the future}, \textit{don't condition on having reached an absorbing state} and \textit{stick to the real world}. 
These tenets are recurring themes in this paper.
To deal with the practicalities of actual trials we use models that represent the  disease process as well as features such as intercurrent events. Responses and features refer to  particular aspects of a process that are of key interest. For example, the three-state diagram of Figure \ref{multistateDiagram}(a) may represent possible disease paths, but in a study aiming to reduce the occurrence of the intermediate event the time of entry to state 1 may be the response of interest. Figure 2 portrays a much more complicated setting where intercurrent events may occur.

We begin with an idealized setting in which individuals in a trial experience illness-death processes, with no premature dropouts before a common end of followup time $C$. The first step in specifying an estimand involves identifying an observable process $\{Z(t), t > 0\}$ and specifying a process feature on which treatment comparisons will be based. 
This can be surprisingly challenging in complex settings and will depend on the clinical meaning of the different states and the primary goals for the experimental intervention. In randomized trials of palliative therapies treatments might aim to reduce the occurrence of an adverse non-fatal event; for example \cite{Cook2009} and Section 1.2.1 above describe  trials involving the reduction of fractures and other skeletal events in patients with advanced breast cancer. In this case the response of interest could be entry to state 1, with the associated feature  the $0 \rightarrow 1$ intensity function $\lambda_{01}(t)$  or the cumulative incidence function $F_1(t)$. In oncology trials aiming to prolong survival in patients with advanced cancer, state 1 might represent disease progression, and  the overall survival time $T_2$ or disease-free survival time $T$ is often considered as the response of interest \citep{Rittmeyer2017, Powles2018}. 
Once a  process feature has been identified, the second step is the specification  of an estimand $\beta$.  
One option is to take a specific time $\tau$ and a feature such as $F_1(\tau|X)$, and to  define $\beta$ as the ratio or difference in 
$F_1(\tau|X=1)$ and $F_1(\tau|X=0)$.
This requires minimal assumptions, but estimands will naturally vary according to the chosen value of $\tau$ which may be contentious.
If the process feature is a function of time such as $F_1(t|X)$ or $\lambda_{01}(t|X)$, a second option is to adopt  modeling assumptions that provide a one-dimensional estimand. In the case of entry to state 1, an estimand  $\beta$ is frequently defined by assuming $F_1(t|X=1)=\exp(\beta) F_1(t|X=0)$ or $\lambda_{01}(t|X=1)=\exp(\beta) \lambda_{01}(t|X=0)$. 
This is not strictly necessary since we could, for example, define $\beta$  as the maximum of  $|F_1(t|X=1)$ - $F_1(t|X=0)|$  over $(0,\tau]$.  
However, power calculations for tests of no treatment effect used in planning studies are most conveniently formulated in terms of model-based estimands, and we will focus on them. 
\newpage
Process features and related estimands can be distinguished according to whether they do or do not condition on previous process history. We refer to the former as dynamic features; process intensity functions are of this type.  We term features that do not condition on previous history as marginal features; the cumulative incidence function $F_1(t|X)$ is an example.  We will also refer to features as having 
a \textit{marginal} interpretation or 
a \textit{dynamic} interpretation.
Causal inference for treatment effects based on marginal features is in principle straightforward: it is facilitated 
by the random allocation of treatment to trial participants at time $t=0$.
Transition intensities are defined conditional on the process history and estimands such as intensity ratios are geared towards process dynamics \citep{Aalen2012}. Intensity functions are crucial to a full understanding of a disease process, and to an understanding of specific marginal features,  but 
for reasons we expand on below,  are not suited for simple causal inference based on randomization.
They are, however, important for a deeper understanding of causal mechanisms which in processes must address time-varying factors and random events.
For example, in view of the relationship (\ref{CIF}), the effects of $X$ on both process intensities $\lambda_{01}(t|X)$ and $\lambda_{02}(t|X)$ must be 
considered in order to understand what has produced an observed effect in $F_1(t|X=0)$ and $F_1(t|X=1)$. {Chapter 9 in \cite{Aalen2008} provides a thoughtful survey of aspects and approaches to causality in the context of event history processes.}

We support the common position that primary assessment of treatment effects in randomized trials should be based on marginal features and estimands.  Beyond this, we argue that they should possess three fundamental properties:

\begin{enumerate}
\item
An estimand should represent the difference between treatment groups with respect to a clinically relevant marginal feature.

\item
Features and estimands should be interpretable in the real world, meaning the response involved should be an element of the observable process.

\item
Estimands should not be  sensitive to uncheckable assumptions; 
models on which an estimand is based should be assessed using available data.
\end{enumerate}

The \textit{first principle} ensures that any findings will have clear relevance for treatment decisions. 
While this appears a straightforward objective it can be challenging to satisfy this principle in the face of intercurrent events.
The \textit{second principle}, that estimands be interpretable "in the real world", may seem self-evident but estimands that do not satisfy this condition are often adopted. For a competing risks process the subdistribution hazard function (see \citealp{Fine1999}) corresponding to $F_1(t|X)$ is a well-known functional that violates our second principle.
Individuals who have previously experienced event-free death are included in the risk set at time $t$ when defining and estimating the sub-distribution hazard but 
such individuals are not genuinely at risk of the non-fatal event in the real world \citep{Andersen2012.1, Putter2020}. 
Moreover, most of the recent work in causal inference defines estimands based on a potential outcome for each  individual subject, corresponding to each of the treatments under study; in the real world of most trials, 
only one of the treatments is received by an individual, making the other potential outcomes "counterfactual".  In this case  inter-individual treatment effects such as the difference in the two potential outcomes are not observable in the real world. 
Many proposed solutions to challenges with intercurrent events also fail to satisfy principle 2: they involve specification of higher dimensional potential outcomes, taking us further away 
from the real world.
Estimands of this type include the survivor average causal effect and other estimands in \cite{Comment2019}, \cite{Stensrud2020}, \cite{Xu2020} and \cite{Young2020}. The \textit{third principle} ensures that one can assess the adequacy of an underlying model and related assumptions, and their effect on the validity of conclusions.
Of course the more complex the disease and disease management processes are, the more difficult it is to select a single  primary estimand that adequately summarizes treatment effects.
Secondary analyses that enhance understanding of intervention effects and more thoroughly characterize the overall response to treatment are an important aspect of clinical trials.
}

\subsection{Problems with conditioning on features of the process history}
\label{sec2.3}
In Sections \ref{sec3} and \ref{sec4} we discuss marginal estimands for specific processes, but first we  review problems with causal interpretations of intensity functions and other conditional process features. 
In particular, estimands such as intensity ratios for experimental versus control treatment groups do not have a simple causal interpretation.  
Although treatment $X$ is randomly assigned at time $t=0$ and is thus independent of other covariates $V$ affecting the disease process, at a later time $t$ the 
random allocation of $X$ and independence of $V$ no longer holds among those at risk for the event in question. 
Confounding induced by conditioning on post-randomization events that may also be responsive to the treatment is a well-known phenomenon; 
it has been comprehensively discussed by \cite{Hernan2010}, \cite{Aalen2015} and \cite{Martinussen2020} for the standard survival case, where the hazard function at time $t$ conditions on being alive then. In illness-death processes, time-dependent confounding is more involved; for example the intensity function for the non-fatal  event conditions on being alive and event-free, and confounding depends on the effect of $V$ on both event types. Hazard or intensity ratios nevertheless remain popular in the analysis of randomized trials, so we illustrate here the impact of time-related confounding in an illness-death process. 

For simplicity, we let $V$ be a binary covariate with $P(V=1)=0.5$ and independent of $X$ due to randomization ($X \perp V$). 
We let $P(X=1)=P(X=0)=0.5$ and assume that the true illness-death process has intensity functions $\lambda_{0k}(t | X,  V) = \lambda_{k} \exp(\gamma_{kx}X + \gamma_{kv}V)$ for $k=1,2$; 
thus $V$ may affect the intensity for both the non-fatal event and death, and both intensities are of proportional hazards form.  
The conditional joint distribution of $X$ and $V$ for subjects in state 0 at time $t^-$ is
$$P(X, V | Z(t^-)=0) = \dfrac{P(Z(t^-)=0 | X,  V)P(X)P(V)}{\mathbb{E}_{XV}(P(Z(t^-)=0 | X, V))}$$ and this cannot be factored into separate functions
of $X$ and $V$, so $X \not \perp V | Z(t^-)=0$.  
In addition, the marginal intensity function 'averaged' over $V$ has the form 
\begin{align*}
 \lambda_{01}(t | X=x, Z(t^-)=0) &= \mathbb{E}_{V}(\lambda_{01}(t | X=x, V) | X,  Z(t^-)=0) \\
 &= \lambda_{1}\exp(\gamma_{1x}x) \mathbb{E}_{V}(\exp(\gamma_{1v}V) | X=x, Z(t^-)=0),
\end{align*}
where $P(V | X=x, Z(t^-)=0) = P(Z(t^-)=0 | X=x, V)P(V)/\mathbb{E}_V(P(Z(t^-)=0 | X=x, V))$. 
As a result the marginal intensity ratio $IR_{01}(t)=\lambda_{01}(t | X=1, Z(t^-)=0)/\lambda_{01}(t | X=0, Z(t^-)=0)$ is given by
\begin{equation} \label {marg-int-ratio}
\exp(\gamma_{1x}) \biggl[  \dfrac{\exp(\gamma_{1v})P(V=1 | X=1, Z(t^-)=0) + 
   P(V=0 | X=1, Z(t^-)=0)}{\exp(\gamma_{1v})P(V=1 | X=0, Z(t^-)=0) + P(V=0 | X=0, Z(t^-)=0)} \biggr].
\end{equation}
If $\gamma_{1v} \neq 0$ this is a function of time $t$ so the intensity ratio for experimental versus control subjects  is not a constant.  
Consequently there are two key points: 
(i) the marginal intensity ratio at a given time $t>0$, $IR_{01}(t)$, cannot be interpreted causally if confounders $V$ are omitted, and 
(ii) if the true intensities conditional on both $X$ and $V$ are proportional, so that  there is a scalar treatment effect $\gamma_{1x}$ that applies at all times $t$,  
the marginal intensity ratio is time-dependent so does not yield a scalar estimand. 

To illustrate this numerically  we set $\gamma_{1x}=\log(0.75),  \gamma_{2x}=\log(0.9)$ and considered three values $0,  \log(2.0)$ and $\log(3.0)$ for each of 
$\gamma_{1v}$ and $\gamma_{2v}$.   
We chose the administrative censoring time to be $C=1$ and for each set of regression coefficients determined $\lambda_1$ and $\lambda_2$ so that $P(T \leq 1)=0.6$
and so that the conditional probability of entry to state $1$ given $T \leq 1$ was either $0.05, 0.4, 0.6, 0.8$ or $1.0$.  
We incorporated loss to follow-up by introducing an independent random right-censoring time $R$ which followed an exponential distribution with rate $\rho$,  with $\rho$ set to satisfy $\pi_R=P(R < T_1 | T_1 < \min(T_2, 1))=0.2$. 
Thus 20\% of the non-fatal events occurring before the administrative censoring time $C$ are censored due to early withdrawal.
As is common in primary analyses of clinical trials, we consider a  cause-specific hazards model for the non-fatal event of the proportional hazards form 
$\lambda_{01}(t | X) = \tilde{\lambda}_{1}(t)\exp(\phi_1 X)$ where $\phi_1$ is estimated by maximizing a Cox partial likelihood.   As we describe below,  the true hazard ratio varies with time $t$ so this model is misspecified. However, 
the maximum likelihood estimator $\hat{\phi}_1$ converges in probability in large samples to a limit $\phi^{*}_1$ and we first consider these values; see the Appendix \ref{app-Cox} for details on how $\phi^{*}_1$ can be obtained.

\begin{sidewaysfigure}
\scalebox{0.59}{ 
\includegraphics{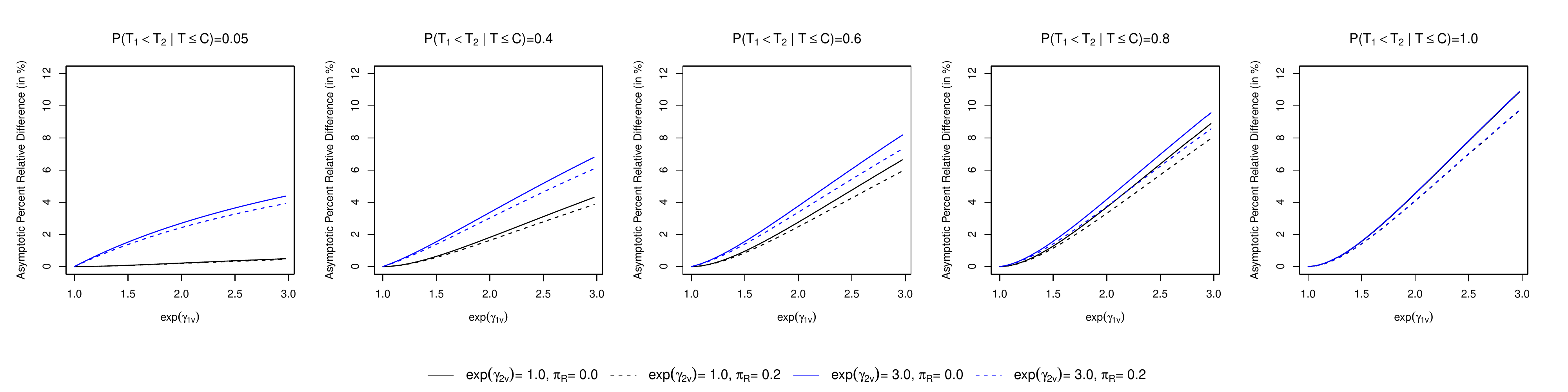}
}
\scalebox{0.59}{ 
\includegraphics{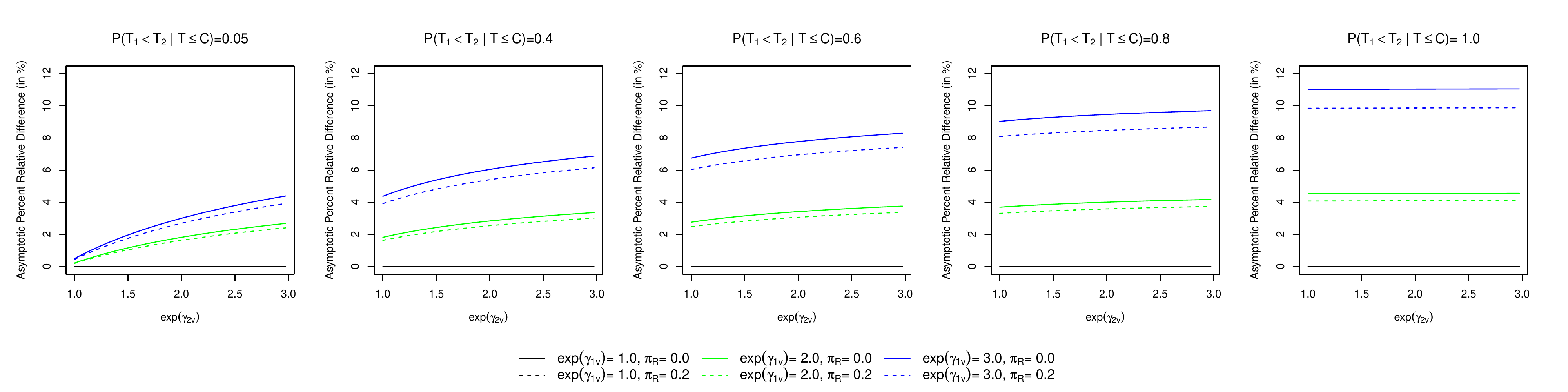}
}
\caption{Asymptotic percent relative difference $100(\phi^{*}_1 - \gamma_{1x})/\gamma_{1x}$ where $\phi^{*}_1$ is the limiting value of the estimator from a cause-specific Cox model for the $0 \longrightarrow 1$ transition when the covariate $V$ is omitted; $C=1, P(X=1)=0.5, P(V=1)=0.5,P(T \leq C)=0.6,  \gamma_{1x}=\log(0.75), \gamma_{2x}=\log(0.9).$}
\label{BiasCoxIntensityAnalysis}
\end{sidewaysfigure}

Figure $\ref{BiasCoxIntensityAnalysis}$ plots $100(\phi_1^{*}-\gamma_{1x})/\gamma_{1x}$,
the percent relative difference between $\phi^{*}_1$ and $\gamma_{1x}$, the effect of $X$ in the true process conditional on $X$ and $V$, 
as a function of $\exp(\gamma_{1v})$ (upper panel) and $\exp(\gamma_{2v})$ (lower panel). 
When $V$ affects the fatal event only (i.e.  $\gamma_{1v}=0$), $\phi^{*}_1$ equals the conditional treatment effect $\gamma_{1x}$ in the true process; 
see (\ref{marg-int-ratio}) and the bottom set of panels in Figure \ref{BiasCoxIntensityAnalysis}.
For all other scenarios, the limiting value $\phi^{*}_1$ differs from $\gamma_{1x}$. 
In settings where the non-fatal event occurs most often (right hand panels, bottom row),  the relative difference is about $10 \%$.
The magnitude of the difference increases with stronger effects of $V$ on the fatal and non-fatal event intensities.  
The relative difference depends on the probability of being event-free at time $t^-$, which is given by $(\ref{EFS})$, and on  the proportion of non-fatal events. 
In the extreme case of $P(T_1 < T_2 | T \leq 1) =1$,  $\lambda_{02}(t|X,V)$ is zero; the competing risk setting coincides with the standard survival 
case then and our results align with the points made in \cite{Aalen2015}. 
If on the other hand the non-fatal event happens rarely,  the relative difference is smaller but its dependence on $\gamma_{2v}$ becomes stronger. 
As noted by \cite{Struthers1986}, we also find that  $\phi_{1}^{*}$ depends on the censoring distribution.

\begin{figure}
\centering 
\scalebox{0.65}{ 
\includegraphics{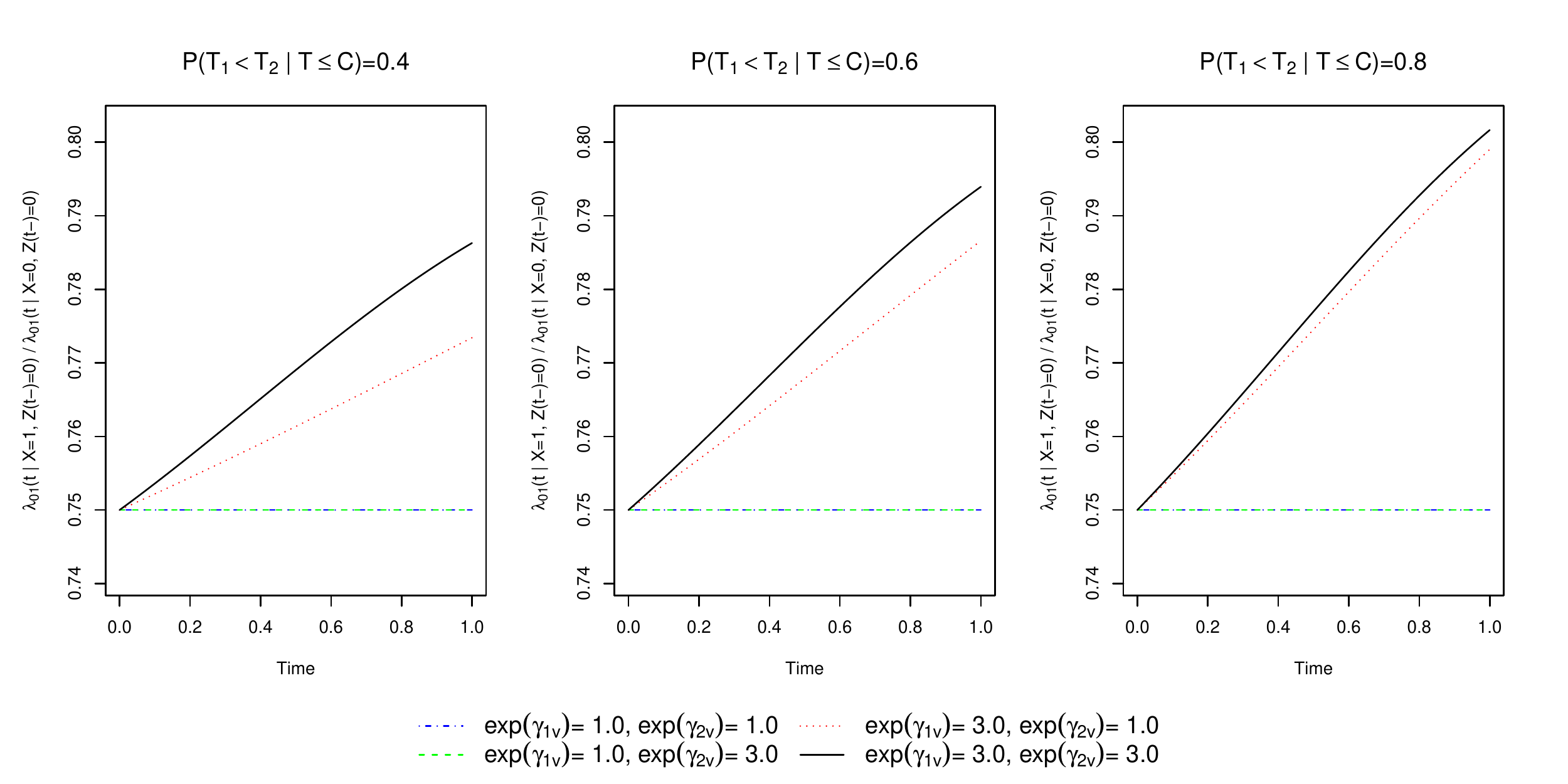}
}
\caption{Plots of $IR_{01}(t)$ when the cause-specific hazards functions depend on $X$ and $V$; $C=1, P(X=1)=0.5, P(V=1)=0.5,P(T \leq C)=0.6,  \gamma_{1x}=\log(0.75), \gamma_{2x}=\log(0.9)$.}
\label{MarginalHR}
\end{figure}

Figure $\ref{MarginalHR}$ depicts the true marginal intensity ratio $IR_{01}(t)$ over the time interval $[0,1]$ for different values of $\gamma_{1v}, \gamma_{2v}$ and 
$P(T_1 < T_2 | T \leq 1)$.  
As noted earlier, unless $\gamma_{1v}=0$, $IR_{01}(t)$
varies with time so that the marginal proportional hazards model is misspecified; this further complicates the interpretation of the effective estimand $\phi^{*}_1$. 
The magnitude of variation in the intensity ratio over time depends on the effects of $V$ on the intensity functions for the non-fatal and fatal event; 
the effects on $\phi^{*}_1$ seen in Figure \ref{BiasCoxIntensityAnalysis} are related to this.  

To summarize,  intensity-based comparison of treatment groups is not recommended for the specification of estimands and primary analysis. Even if models adequately describe the observed data, intensities condition on post-randomization events and do not permit randomization-based causal interpretations.  In addition, the almost inevitable presence of other factors $V$ that affect the disease process further complicates a causal interpretation of treatment effects.  For more comprehensive marginal models or  intensity-based  models  used for secondary analysis,  known baseline covariates $V$ can be included, with inferences about the effect of $X$ now adjusted for $V$.   Even then there may  exist unknown or unobserved covariates that affect the process, so  some caution should be exercised when interpreting treatment effects.

\subsection{General frameworks for expressing marginal treatment effects}
\label{GeneralEstimandFramework}

We now consider frameworks for specification of marginal effects and discuss estimability and model assumptions.  
Most proposed estimands are based on the distribution of the time to some  event; for an illness-death process, 
the times $T_1, T_2$ and $T$  are all  used in certain settings.  
Quantiles or expected values of these times in "restricted" form are also used; 
 the restricted mean time event-free over a specified time interval $(0,\tau]$ is $RMT(\tau)=\mathbb{E}(\min(T,\tau))$.
Estimation of state occupancy probabilities by treatment group, $P(Z(t)=k|X)$, is also appealing in many cases and can  be related to distributions of
 event times.  
For example, $P(T>t|X)=P(Z(t)=0|X)$ and $RMT(\tau|X)= \int_{0}^{\tau} P(Z(u)=0|X)du$.  
Similarly, in an illness-death process $P(T_2 > t|X)=P(Z(t) \in \{0, 1\}|X)$.  
A third related framework is  utility-based: we assign utility scores to each state,  and then consider  the average cumulative utility over a specified time period. 
This has been used in quality of life comparisons in oncology trials \citep{Gelber1989, Gelber1995, Glasziou1990, Cook2003}.
We will briefly comment on utility-based estimands in Section \ref{sec5.2}. 

In the time to event framework  $P(T _E \le t)=F_E(t)$ denotes the distribution function for the time $T_E$ to a defined event $E$.
With a specified time horizon, one can estimate $F_E(\tau|X)= P(T_E \le \tau|X)$ separately for each treatment arm and compare them; this can be done nonparametrically.
A more common approach for specifying an estimand $\beta$ is through transformation models of the form 
\begin{equation}
\label{EventTime}
g(F_E(t|X)) = \alpha_{0}(t) + X \beta, 
\end{equation}
where $g$ is a known  differentiable  monotonic function on $(0,1)$ and $\alpha_{0}(t)=g(F_E(t|X=0))$ is a monotonic  
function with $\alpha_0(t) \downarrow - \infty$ as $t \downarrow 0$ \citep{Scheike2007, Scheike2008}.  
Such a model makes a strong assumption about the form of any difference in $F_{E}(t|X=1)$ and $F_{E}(t|X=0)$ and should be checked in a given setting.  
We note that the common practice of assuming a Cox proportional hazards model for times $T_E$ produces a model of the form (\ref{EventTime}); as discussed, we argue that the treatment effect $\beta$ in such a model should be interpreted for causal purposes in terms of (\ref{EventTime}) and not as a hazard ratio.
We discuss special cases of model (\ref{EventTime}) in the context of  a competing risk process in Section \ref{sec3}.   


Finally,  tests of no treatment effect are a key component of primary analysis,  and power calculations are important in planning trials.  Hypothesis tests can be 
based on nonparametric estimates of process features but to address power we usually want  a parametric assumption that gives an ordering of alternative hypotheses.  
We assume in further discussion that  hypotheses can be expressed in terms of a parametric estimand $\beta$.   

\section{Illness-death and competing risk processes} \label{sec3}

\subsection{Estimands based on marginal process features}
We now take a closer look at estimands based on marginal features of a process. In this section we consider illness-death and competing risks processes,  given their wide applicability. More general multistate processes involving several intermediate health states, recurrent events or reversible conditions are considered in the next section. Additional issues complicating the interpretation and selection of estimands arise when post-randomization events involving compliance, treatment switching, rescue therapy, or loss to followup occur; these are also discussed in Section 4.

We first  consider illness-death settings where times $T_1$ or $T$ are responses of interest. In their case it is sufficient to focus on the associated competing risks process $\{Z(t),  t \geq 0 \}$ in Figure \ref{multistateDiagram}(b), where $1 \rightarrow 2$  transitions are not considered. The exit time from state $0$ is $T = \inf\{t>0:Z(t)\neq 0 \}$ and the  indicator $\varepsilon = Z(T) \in \{1, 2 \}$ records the 
type of event. We let  $X$ and $V$ denote treatment indicator and other covariates, as before.  The intensity functions in this case are often called cause-specific hazard functions:
$$ \lim_{\Delta t \downarrow 0}  \dfrac{P(T \in [t, t+ \Delta t), \varepsilon=k |  T \geq t, X,V)}{\Delta t} = \lambda_{0k}(t | X,V), \ \ \ \ \ \ \ \ k=1,2 $$ 
and they completely determine the competing risks process. Models for the intensities  have to be chosen, 
but methodology and software for fitting and checking models is widely available \citep{Cook2018}. The event time $T$ has survivor function $S(t|X,V)$ given by (\ref{EFS})
and the cumulative incidence function for $k=1$ is given by (\ref{CIF}); each
depends on both cause-specific hazard functions. 

To illustrate the specification of estimands based on marginal features,  we consider  quantifying the difference between $F_1(t|X=1)$ and $F_1(t|X=0)$; for 
convenience we continue to denote the cumulative incidence function as $F_1$, though we condition now on $X$ alone.
One approach is to estimate the two functions nonparametrically 
\citep[Section 3.2]{Cook2018}
and then to use some one-dimensional measure of their difference. \cite{Zhang2008} proposed comparison at a specified time point $\tau > 0$.  
Such estimands do not provide a full picture of the effect of treatment on $F_1(t)$, nor do estimands such as differences in restricted mean time or 
the maximum of $|F_1(t|X=1)-F_1(t|X=0)|$ over a time period $(0,\tau]$.  

A second way to obtain a one-dimensional estimand is by making parametric assumptions about the difference between $F_1(t|X=1)$ and $F_1(t|X=0)$.  
The most common approach has been to use  models that are of generalized linear form,  as in (\ref{EventTime}).
The  function $\alpha_{0}(t)=g(F_1(t|X=0))$ may be modelled parametrically or nonparametrically and the regression 
coefficient $\beta= g(F_{1}(t \ | \ X=1))-g(F_{1}(t \ | \ X=0))$ is an estimand. 
Several functions $g(u)$ have been considered in the literature \citep{Fine1999, Gerds2012}; common choices are  
$g(u)=\log(u)$, ${\rm logit}(u)$ and $\log(-\log(1-u))$, often called the complementary log-log, or cloglog  transform.  
  Assumed models should of course adequately represent treatment group differences; if this is not the case the interpretation of estimates $\hat{\beta}$ is 
affected, as we illustrate in the next section. 

Parametric or semi-parametric estimation for arbitrary transformation models for $F_1(t|X)$  can be based on so-called direct binomial estimating 
functions \citep{Scheike2007} or on pseudo-value methods \citep{Klein2005}.  
These methods avoid modeling $F_2(t|X)$, whereas maximum likelihood estimation requires such a model when some individuals are still in state $0$ at the end of 
followup. 
The most widely used model is based on the $cloglog$ function; 
it can be used in analysis based on a weighted partial likelihood estimating function targeting the hazard function for the sub-distribution $F_1(t|X)$ \citep{Fine1999}. 
As noted previously, this "hazard" function does not have a real world interpretation \citep{Andersen2012.1}; however, although the associated estimand $\beta$ 
cannot be interpreted as a hazard ratio, it can be interpreted  in terms of the $cloglog$ transformation model,  and the estimation procedure of \cite{Fine1999} is valid in this context. These methods are reviewed by \citet[Section 4.1]{Cook2018}, who provide references to software. In addition to packages mentioned there, the R-functions coxph and cifreg now handle the Fine-Gray method.

The adequacy of a transformation model for $F_1(t|X)$ can be checked in a given setting by plotting $g$-transformed  nonparametric Aalen-Johansen estimates for each treatment group.  
We have found that the $cloglog$ model represents the results quite well for many oncology trials in which state $1$ represents disease progression or the onset
of complications.  One can explore different transformations through parametric families  of
functions  $g(t, \nu_k)$ specified by an unknown paramter $\nu_{k}$; this is easier if parametric assumptions are also made about the function $\alpha_0(t)$. \cite{Jeong2007} considered a generalized Burr form 
\begin{equation}
\label{parametricCIF}
F_{k}(t \ | \ X) =  1 - \biggl(1+\dfrac{\exp(\alpha_{0k}(t,  \phi_k))\exp(X\beta_k)}{\nu_k}\biggr)^{-\nu_k}. 
\end{equation}
for $k=1,2$ and developed maximum likelihood procedures.  Model $(\ref{parametricCIF})$ includes  parametric versions of the $logit$ and $cloglog$ transformation 
models ($\nu_{k}=1$ and $\nu_{k} \longrightarrow \infty$ respectively) as special cases.   
Since $F_k(t|X) < 1$ we need $\alpha_{0k}(t)$ to approach a finite limit as $t$  increases; $\beta_k$ must also be constrained to keep $F_k(t|X=1) < 1$.  
In addition $F_1(t|X) + F_2(t|X)$ cannot exceed one, which may require additional constraints.  
In practice  a transformation model should provide an adequate representation up to some maximum followup time $\tau$  and these constraints may not have much 
impact in some settings.  
Other families of models have also been proposed,  some of which automatically constrain the cumulative incidence functions to sum to 
one (e.g. \citealp{Gerds2012}).

For some illness-death settings  the time $T_2$ of death or failure may be the preferred feature, regardless of whether or not an individual passes through state 1.  
Marginal models for $F_D(t|X)=P(T_2 \le t|X)$ such as  (\ref{EventTime}) can be applied,  and  permit descriptive causal interpretations of treatment effects.  
Cox proportional hazards models are often fitted for $T_2$ given $X$, with  the regression coefficient $\beta$ for $X$  a common estimand.  
Once again, this should be interpreted in terms of the associated $cloglog$ transformation model which the distribution function $F_D(t|X)$ in the Cox model satisfies; 
the model can be checked as described above.

Finally,  given the importance of estimands based on marginal process features in primary  estimation and testing of treatment effects,
 we recommend that the study planning stage include careful consideration of process intensities, and  how treatment may affect them.  
This promotes an understanding of what types of marginal features and models to employ and informs decisions concerning sample size and duration of followup.  
In the next section we provide some numerical illustrations of the connection between intensities and marginal features, and 
Section \ref{sec4} contains illustrations involving more complex settings with intercurrent events.

\subsection{Illustrative calculations for cumulative incidence function regression}\label{sec3.2}

To provide insight into the impact of process intensities on marginal process features, we consider some illustrative calculations.
A pragmatic point of view is that models only approximate reality, and when we define an estimand $\beta$ based on assumptions about a process (that is, a model), along with a method of estimating it,  the true  estimand is actually the limiting value $\beta^*$ of the estimator $\widehat{\beta}$ as the number of subjects $n$ becomes arbitrarily large. The value $\beta^*$ is sometimes referred to as the least false parameter \citep{Grambauer2010}, 
but more precisely it is the parameter value for which the assumed model family is "closest" (in the expected log likelihood or Kullback-Leibler sense) to the true process or distribution in question. Equivalently, it is the value for which the estimating function used to obtain the estimator of $\beta$ has expectation zero with respect to the true data generation process.

We present a simple illustration for the illness-death setting, with the true process assumed to have  intensity functions 
$\lambda_{0k}(t|X)= \lambda_k \exp(\gamma_k X)$ for $k=1,2$.
We focus on the cumulative incidence function $F_1(t|X)$ for the non-fatal event,  and consider transformation models (\ref{EventTime}) 
where $g(u)= \log(u)$ or $\log(-\log(1-u))$.  
The interpretation (and relevance) of $\beta$ depends on whether the transformation model adequately approximates the difference in $F_1(t|X=0)$ and $F_1(t|X=1)$.  
We can also ask how $\beta^*$ is related to the baseline intensities and treatment effects in the true process. We let $P(X=1)=P(X=0)=0.5$ and considered true processes with $\gamma_1=\log(0.75)$ and $\gamma_2=\log(0.9,1.0,1.1)$; thus the treatment reduces the intensity for the non-fatal event by one quarter and gives either a mild decrease, no change, or a mild increase in the intensity for the fatal event.  
We took the administrative censoring time to be $C=1$ and for each value for $\gamma_2$, set $\lambda_1$ and $\lambda_2$ so that $P(T \leq 1)$ was either 0.2 or 0.6 and so that the probability of entry to state 1,  given $T \le 1$, was either 0.4, 0.6 or 0.8. We note that the assumed form for the intensities is for illustration; as we saw in Section \ref{sec2.3}, the presence of other covariates $V$ that affect intensities can produce non-proportional intensities when only $X$ is modeled.

\begin{figure}
\centering
\scalebox{0.70}{
\includegraphics{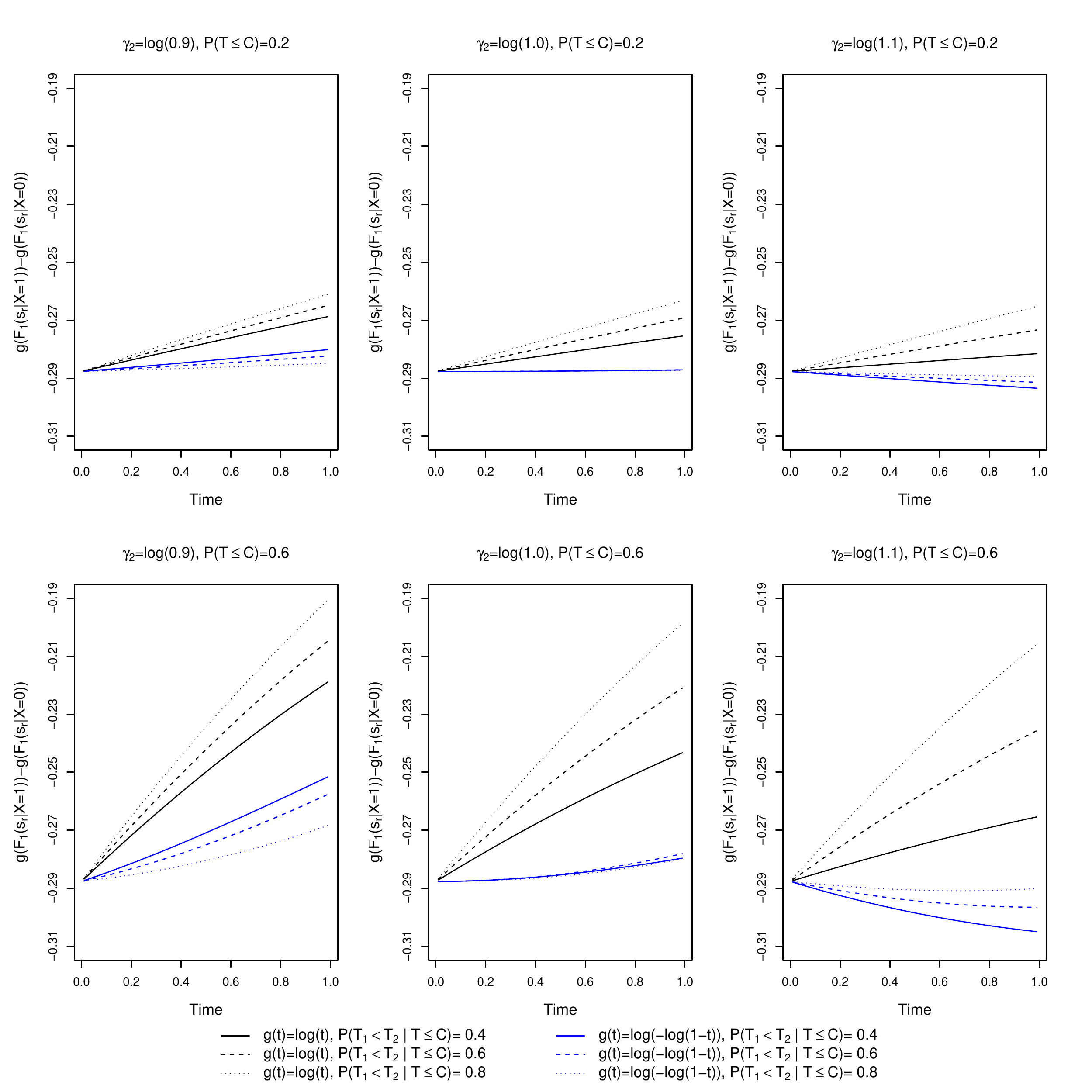}
}
\caption{Plots of $g(F_1(s_r | X=1))-g(F_1(s_r | X=0))$ for $20$ equi-spaced values of $s_r$ in $(0,1)$ and different values of $P(T \leq C),  P(T_1 < T_2 | T \leq C)$ and $\gamma_2$ under the $log$ and $cloglog$ transformation models; $C=1, P(X=1)=0.5, \gamma_1=\log(0.75)=-0.288$.}
\label{PlotAdequacy}
\end{figure}

We consider first the adequacy of the transformation models by computing the true values of $F_1(s_r|X)$ for 20 equi-spaced values of $s_r$ in (0,1) and $X=0,1$. 
For each transformation, we  calculated $g(F_1(s_r|X=1))-g(F_1(s_r|X=0))$ for $r=1, \ldots,20$ and display the results in Figure \ref{PlotAdequacy}.  
Departures from a constant line are indicative of model misspecification.  
In most cases the $cloglog$ transformation model is a fairly good approximation, especially when the treatment has no effect on the fatal event, whereas the $\log$ 
transformation model approximates the differences in $F_1(t | X)$ less well.  
It can be shown that at $t=0$ the difference in the transformed functions is $\gamma_1= \log(0.75)= - 0.288$ under either model.  
For $g(u)=\log(-\log(1-u))$,  adequacy of the transformation model varies with the probability of entry to state $1$ and the treatment effect $\gamma_2$ 
on the fatal event intensity; the model improves with increasing $P(T_1 < T_2 | T \leq 1)$ and as the magnitude of $\gamma_2$ decreases.  
For $g(u)=\log(u)$,  adequacy of the model likewise varies with the parameter setting.  
We have found that these results persist across various plausible settings, and  we restrict our attention to $g(u)=\log(-\log(1-u))$ in the remainder of this discussion.

We considered  estimation of $\beta$ for the $cloglog$ transformation model using the two main  methods: 
(a) the method of \cite{Fine1999} and 
(b) direct binomial estimation \citep{Scheike2007}. 
Competing risks data under the true process with intensity functions $\lambda_{0k}(t |X) = \lambda_{k}\exp(\gamma_k X)$ as specified above were simulated as follows:  
for each of $X=0,1$ the event time $T | X$ was generated from an exponential distribution with rate 
$\lambda_{01}(t | X) + \lambda_{02}(t | X)$, and given $T=t$,  the event type was drawn from a single binomial trial with non-fatal event probability 
$\lambda_{01}(t | X)/(\lambda_{01}(t | X)+\lambda_{02}(t | X))$.  
As in Section $\ref{sec2.3}$,  we also considered a random withdrawal time $R$, which was generated from an exponential distribution with rate $\rho$.  
We set $\rho$ so that $\pi_R=P(R < T_1 | T_1 < \min(T_2, 1))$ was either $0.0,  0.1$ or $0.2$,  with $\pi_R=0.0$ corresponding to administrative censoring only.  
In scenarios with $\pi_R \neq$ 0.0,  this gave the net censoring time $\min(R, C)$. 
We simulated data sets with $n$ = 1000 observations and for each we obtained estimates of $\beta$ using the Fine-Gray (FG) approach with and without 
weight stabilization, and by direct binomial (DB) estimation based on $6$ equi-spaced time points in $(0,1)$.  
We used the R-functions crr and comp.risk to fit the models,  combined with unstratified Kaplan-Meier estimation to get an estimate of $P(R > t)$.  
Results are shown in Table $\ref{simulation}$ for $N$ = 2000 simulation runs. For each estimation procedure we report the limiting value ($\beta^*$), the mean of the estimates $\hat{\beta}$, their empirical standard error (ESE), the average model-based standard error (ASE) and the empirical coverage probability (ECP) for $\beta^*$ of the nominal $95 \%$ confidence intervals. In Appendix \ref{appendix-FG} and \ref{appendix-DB} we briefly describe how the limiting values $\beta^{*}$ under either estimation procedure can be obtained.


\begin{table}
\centering
\scalebox{0.70}{ 
\begin{tabular}{ c c c cccccc l ccccc}
\hline
\multirow{3}{*}{$P(T \leq C)$} & \multirow{3}{*}{$P(T_1 < T_2 | T \leq C)$} & \multirow{3}{*}{$\pi_R$} & \multicolumn{12}{c}{\textit{\textbf{Method}}}                                                                                                                                                                             \\ \cline{4-15} 
                         &                                            &                          & \multicolumn{6}{c}{\textit{\textbf{Fine-Gray}}}                                                                   & \multirow{11}{*}{} & \multicolumn{5}{c }{\textit{\textbf{Direct Binomial Regression}}}                           \\
                         &                                            &                          & $\beta^{FG*}$ & $\beta_{stab}^{FG*}$ & $\text{mean}(\widehat{\beta}^{FG}_{stab})$ & ESE    & ASE    & ECP  &                    & $\beta^{DB*}$ & $\text{mean}(\widehat{\beta}^{DB})$ & ESE    & ASE    & ECP  \\ \cline{1-9} \cline{11-15} 
\multirow{9}{*}{0.6}     & \multirow{3}{*}{0.4}                       & 0.0                      & -0.2532           & -0.2532                  & -0.2550                                    & 0.1308 & 0.1300 & 95.2 &                    & -0.2702           & -0.2672                             & 0.1437 & 0.1429 & 95.2 \\
                         &                                            & 0.1                      & -0.2532           & -0.2551                  & -0.2579                                    & 0.1341 & 0.1369 & 95.6 &                    & -0.2702           & -0.2693                             & 0.1479 & 0.1486 & 95.1 \\
                         &                                            & 0.2                      & -0.2532           & -0.2572                  & -0.2574                                    & 0.1474 & 0.1451 & 95.0 &                    & -0.2702           & -0.2645                             & 0.1567 & 0.1558 & 95.2 \\ \cline{2-9} \cline{11-15} 
                         & \multirow{3}{*}{0.6}                       & 0.0                      & -0.2601           & -0.2601                  & -0.2594                                    & 0.1067 & 0.1059 & 94.6 &                    & -0.2742           & -0.2704                             & 0.1180 & 0.1161 & 94.4 \\
                         &                                            & 0.1                      & -0.2601           & -0.2617                  & -0.2630                                    & 0.1113 & 0.1116 & 95.1 &                    & -0.2742           & -0.2727                             & 0.1199 & 0.1214 & 95.8 \\
                         &                                            & 0.2                      & -0.2601           & -0.2635                  & -0.2660                                    & 0.1172 & 0.1182 & 95.3 &                    & -0.2742           & -0.2737                             & 0.1278 & 0.1280 & 95.2 \\ \cline{2-9} \cline{11-15} 
                         & \multirow{3}{*}{0.8}                       & 0.0                      & -0.2708           & -0.2708                  & -0.2685                                    & 0.0923 & 0.0917 & 94.6 &                    & -0.2798           & -0.2759                             & 0.1014 & 0.1002 & 95.3 \\
                         &                                            & 0.1                      & -0.2708           & -0.2719                  & -0.2721                                    & 0.0955 & 0.0966 & 95.4 &                    & -0.2798           & -0.2781                             & 0.1045 & 0.1055 & 95.5 \\
                         &                                            & 0.2                      & -0.2708           & -0.2730                  & -0.2681                                    & 0.1020 & 0.1024 & 95.4 &                    & -0.2798           & -0.2715                             & 0.1102 & 0.1119 & 95.6 \\ \hline
\end{tabular}
}
\caption{Asymptotic and empirical properties of Fine-Gray (FG) and direct binomial (DB) regression estimators for $\beta$ in $cloglog$ transformation models; $C=1, P(X=1)=0.5, \gamma_{1}=\log(0.75)= -0.288,  \gamma_{2}=\log(0.9),  n=1000$ subjects,  $N=2000$ simulation runs. Weights in FG and DB approaches are based on covariate-independent
censoring and under the assumption that the censoring model is correctly specified.}
\label{simulation}
\end{table}

Since $\beta^{FG*}, \beta^{FG*}_{stab}$ and $\beta^{DB*}$ solve different expected estimating equations, we do not expect these limiting values to be the same.  However, they are in close agreement, and the size and direction of their departure from $\gamma_1$ is consistent with the departures from a  horizontal line  seen in Figure \ref{PlotAdequacy}.  
In all scenarios, the average estimates under both estimation methods closely approximate the corresponding limiting values, which are close to the value 
$\gamma_1=\log(0.75)$, and the figure provides a simple basis for the interpretation of estimates $\hat{\beta}$, unlike so-called average hazard ratios.  
We also see close agreement between the empirical and  the average model-based standard errors, and that the empirical coverage probabilities are close to the 
nominal level.  Thus for the scenarios considered here,  the $cloglog$ transformation model provides a reasonable approximation for the effect of $X$ on $T_1$, and both  FG and DB estimation of $\beta$ in the $cloglog$  model perform well.

\section{Defining estimands for more complex settings}  \label{sec4}
\subsection{Challenges involving estimands with intercurrent events} \label{sec4.1}

Marginal features, and hence the treatment effect, are usually conceptualized in an idealized setting where individuals under study are compliant, receive medical care according to a prescribed (deterministic or well-characterized stochastic) treatment strategy, and complete followup. Under such circumstances the differences between treatment arms with respect to, say, the proportion of individuals experiencing an event by some time $\tau$ can be attributed to the treatment assigned at randomization.
In practice, events such as termination of followup due to inefficacy or adverse reactions to treatment, the introduction of rescue medication, or treatment discontinuation or switching
may occur over the course of followup,  making the resulting process incompatible with the idealized setting.
Such events are examples of "intercurrent events", defined in the ICH E9 (R1) guidance document \citep{iche9-2017} as
``events occurring after treatment initiation that affect either the interpretation or the existence
of the measurements associated with the clinical question of interest''.
We refer to intercurrent events which preclude observation of the event of interest as type 1 intercurrent events, and intercurrent events which do not preclude their observation but change the
interpretation of the clinical events (due to the condition under which they arise) as type 2 intercurrent events.
A type 1 intercurrent event (IE) can be sub-classified as a type 1A IE if it simply\textit{ impacts the observation} of the clinical event of interest
because of loss to followup or a type 1B IE that \textit{precludes the occurrence} of the clinical events (e.g.  death).
Examples of type 2 intercurrent events are treatment discontinuation without termination of followup, treatment switching, or introduction of rescue therapy that is prohibited by the protocol.
The core challenge with intercurrent events is that  marginal process features differ from those under the idealized setting, making the intended causal analyses challenging.

Consider a clinical trial where  individuals are randomized to receive an experimental treatment ($X=1$) or standard care ($X = 0$).
To simplify discussion we consider a trial involving an illness-death process depicted in
Figure \ref{fig-intensity-based-models}(a),
where states 0, 1 and 2 represent being alive and event-free, alive post-event, and dead, respectively.
We assume individuals begin in state 0 at $t=0$ with followup planned until an administrative censoring time $\tau$.
Let $Z(t)$ denote the state occupied at time $t$ with $Y_k(t)=I(Z(t^-)=k)$, $k=0, 1$, and let $H(t)$ be the process history up to time $t^-$.
The $k \rightarrow l$ transition intensity is denoted by $\lambda_{kl}(t | H(t))$ and we also let
\begin{equation}
\label{qkl}
\lim_{\Delta t \downarrow 0}
\frac{P( Z(t + \Delta t^-) = l | Z(t^-)=k, X)}{\Delta t} =  q_{kl}(t| X)\; ,
\end{equation}
denote the transition rate given $X$ for $(k, l) \in \{ (0,1), (0,2), (1,2) \}$.
The target estimand could be any of those discussed in Sections \ref{sec2} and \ref{sec3} including,
for example,
\begin{itemize}
\item[\textit{i)}] $\beta(\tau) = F_1(\tau | X = 1) - F_1(\tau | X = 0)$ given in Section \ref{sec2.1},
\item[\textit{ii)}] $\beta(\tau) = S(\tau | X = 1) - S(\tau | X = 0)$ where $S(t|X=x) = P(Z(t) < 2|X=x)$
is the survival function for time to death, given $X = x$.
\end{itemize}
These quantities can all be estimated with Aalen-Johansen methods using separate nonparametric estimates of (\ref{qkl}) for each treatment arm.
Estimands based on transformation models of cumulative incidence functions as in (\ref{EventTime}) are also possible but we focus here on those listed above since they require minimal modeling assumptions.

Multistate models such as this can be expanded to incorporate  intercurrent events; this facilitates a clear discussion of marginal features, the interpretation of possible estimands, and offers a framework for
comparing analysis strategies.
Figure \ref{fig-intensity-based-models}(b) depicts an expanded state space for a joint model involving the illness-death process of
Figure \ref{fig-intensity-based-models}(a) and a type 1 intercurrent event that precludes observation of transitions after it.
Figure \ref{fig-intensity-based-models}(c) represents a joint model with a type 2 intercurrent event that does not preclude further followup:
state $0'$ is entered if the intercurrent event occurs in an individual who is event-free, while $1'$ is occupied if an individual is alive but has experienced both the
non-fatal clinical event and the intercurrent event; state $2'$ is entered upon death by individuals who have experienced the intercurrent event.
We let $\{ Z^\circ(s), 0 < s \}$ denote either the five-state process depicted in Figure \ref{fig-intensity-based-models}(b) or the six-state process depicted in Figure \ref{fig-intensity-based-models}(c),  $H^\circ(t)=\{Z^\circ(s), 0< s< t, X\}$ the corresponding process history and define
$Y_k^\circ(t) = I(Z^\circ(t^-)=k)$ for $k=0,1$ or $k=0, 1, 0', 1'$, respectively.
We  let $\lambda_{kl}^\circ(t | H^\circ(t))$ denote the  intensity function for $k \rightarrow l$ transitions within Figure \ref{fig-intensity-based-models}(b) or
Figure \ref{fig-intensity-based-models}(c), with the corresponding rate functions
\begin{equation}
\label{qklcirc}
\lim_{\Delta t \downarrow 0}
\frac{P( Z^\circ(t + \Delta t^-) = l | Z^\circ(t^-)=k, X)}{\Delta t} =  q^\circ_{kl}(t| X)\; ,
\end{equation}
for $(k,l) \in \{ (0,1), (0,2),(0, 0'), (1,2), (1, 1')\}$ or $(k, l) \in \{ (0,1), (0,2), (0, 0'), (1,2), (1, 1'), (0', 1'), (0',2'), (1', 2') \}$.

In the next section we examine three types of scenarios involving intercurrent events:
loss to followup,
discontinuation of the randomized treatment without loss to followup, and
introduction of rescue treatment without loss to followup.
In some clinical trials individuals may switch from the randomized treatment to the treatment of the other arm.
For each setting we will discuss  issues arising from these events,  including possible targets of inference (estimands) and strategies for estimating these quantitites.
We stress the need for careful thought about interpretation of the target estimand and the strength of assumptions required to estimate it.
We focus on estimands relevant to the real world, which invariably leads to analyses within the intention-to-treat (ITT) framework \citep{montori2001}.

\subsection{Some illustrative examples involving intercurrent events} \label{sec4.2}
\subsubsection{Loss to followup } \label{sec4.2.1}

\begin{figure}[!ht]
\begin{center}
\noindent
\begin{minipage}[t]{0.30\textwidth}
\begin{center}
\includegraphics[scale=0.35]{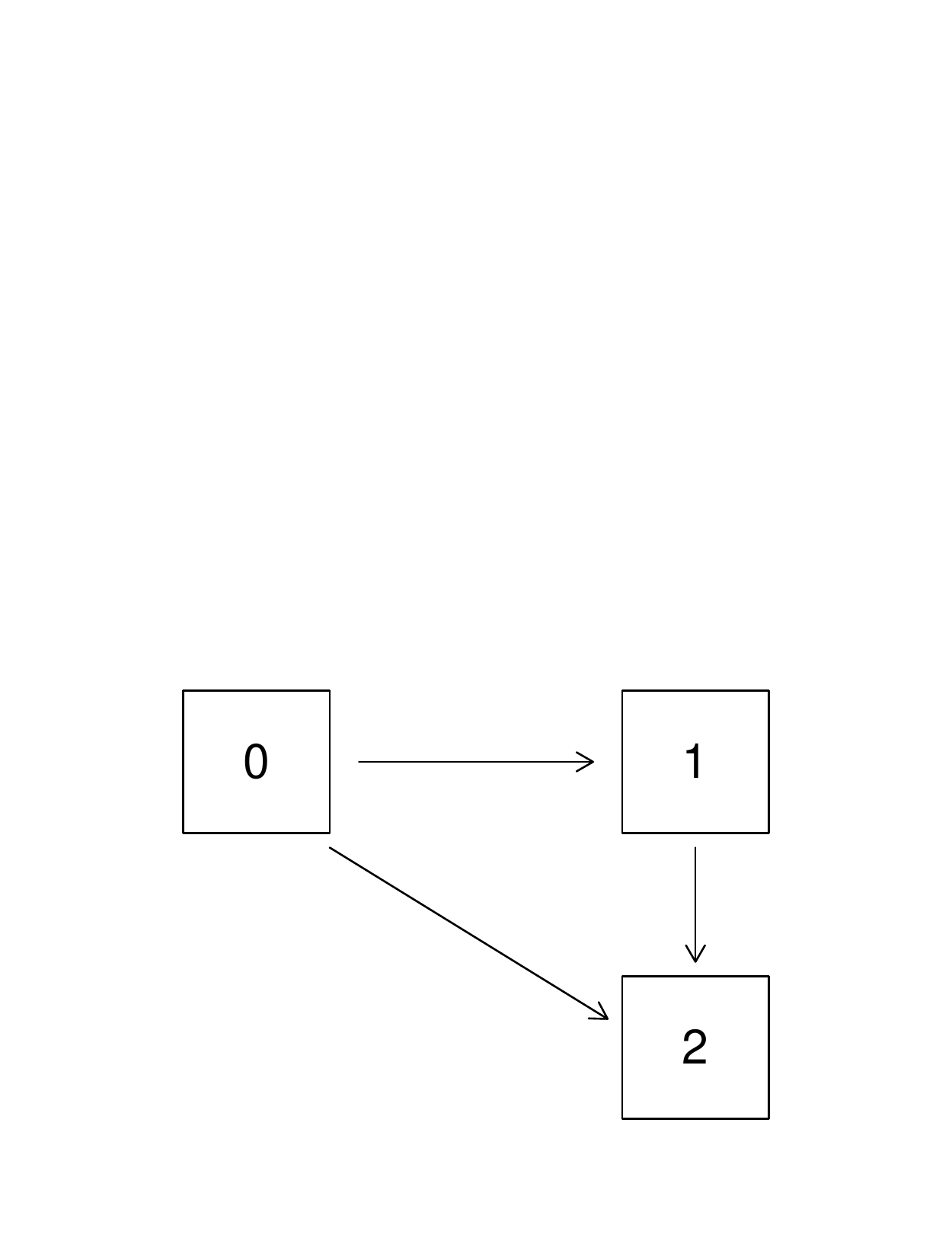} \\
\end{center}
{\footnotesize (a) An illness-death process}
\end{minipage}
\hfill
\noindent
\begin{minipage}[t]{0.30\textwidth}
\begin{center}
\includegraphics[scale=0.35]{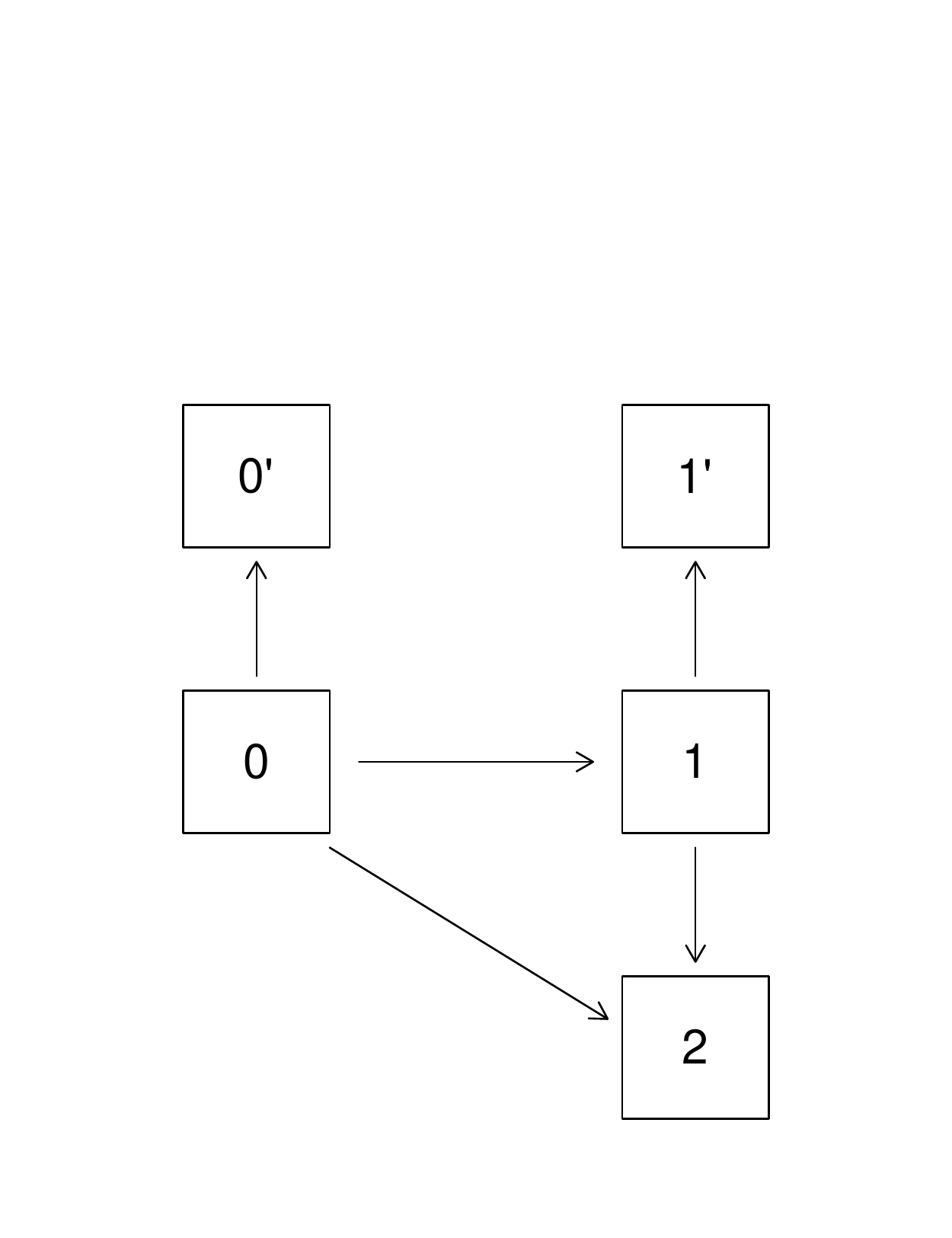} \\
\end{center}
        {\footnotesize (b) A joint illness-death and type 1 intercurrent event process}
\end{minipage}
\hfill
\noindent
\begin{minipage}[t]{0.33\textwidth}
\begin{center}
\includegraphics[scale=0.35]{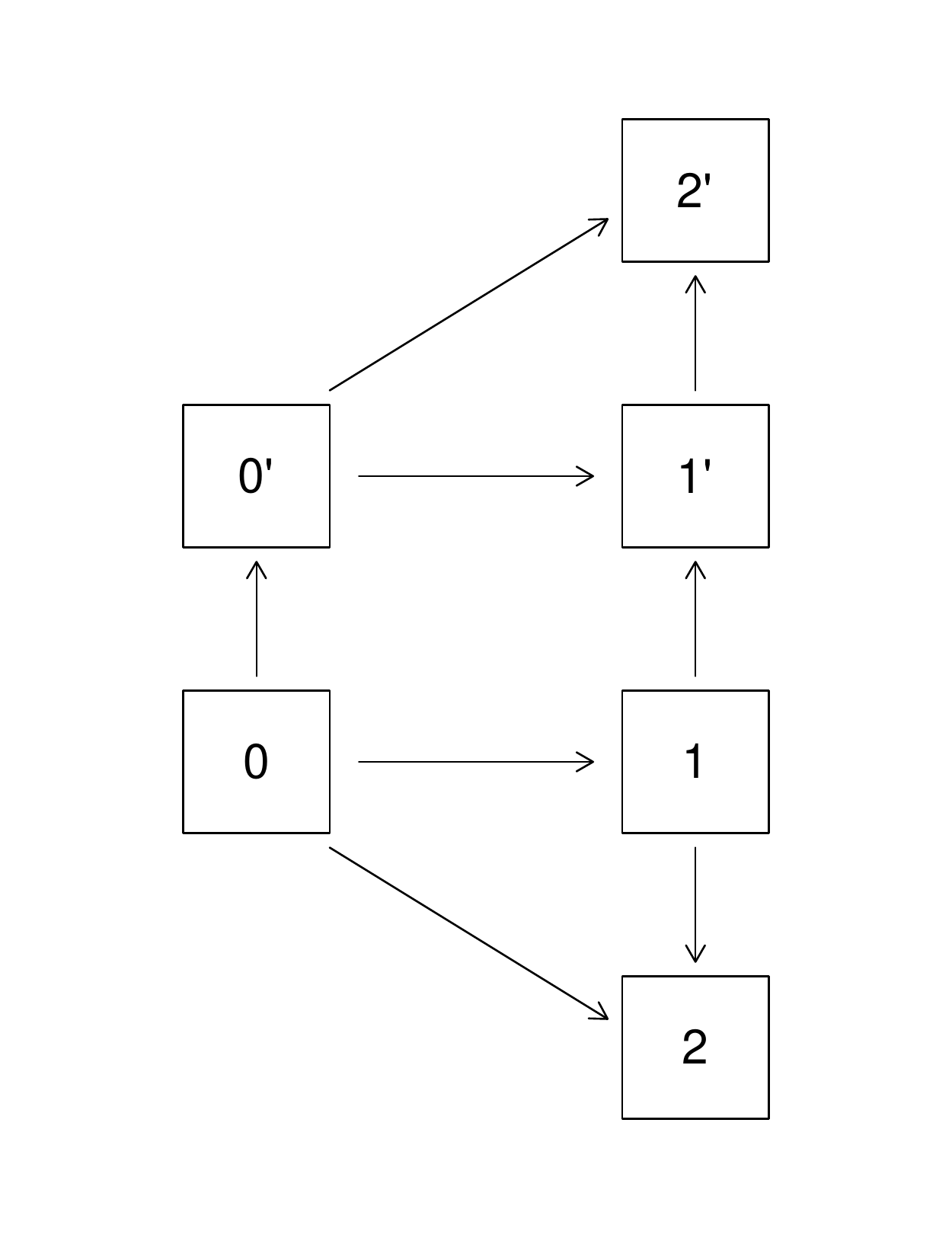} \\
\end{center}
        {\footnotesize (c) A joint illness-death and type 2 intercurrent event process}
\end{minipage}

        \caption{Multistate diagrams for a simple illness-death process (panel (a)) and
                joint models for an illness-death process and type 1 (panel (b)) and type 2 (panel (c)) intercurrent events.}
\label{fig-intensity-based-models}
\end{center}
\end{figure}

Premature loss to followup (LTF) is a type 1 IE (or results from a type 1 IE) and Figure \ref{fig-intensity-based-models}(b) depicts the illness-death  process with states added to represent termination due to LTF.  For a more thorough analysis we need to also consider  the joint model of Figure \ref{fig-intensity-based-models}(c). In this framework, 
\cite{Lawless2019} define conditionally independent LTF for multistate processes observed in cohort studies. The condition in their equation (4a) can be restated in the present notation as
\begin{equation}
\label{ind1}
\lim_{\Delta t \downarrow 0}
\frac{P( Z^\circ(t + \Delta t^-) = l | Z^\circ(t^-)=k, H^\circ(t))}{\Delta t}
=  \lambda_{kl}(t| H(t))\; ,
\end{equation}
for $(k,l) \in \{ (0, 1), (0, 2), (1, 2)\}$.
This condition states that the transition intensities between states $0, 1$ and $2$ are the same for the process under observation as they are for the process of interest depicted in Figure \ref{fig-intensity-based-models}(a).
If baseline covariates or marker processes governing the $j \rightarrow j'$ transitions are present, the intensities for $j \rightarrow j'$ transitions can be modeled to give insights into the types of individuals becoming lost to followup.
These intensities can in turn be used for the construction of inverse probability of censoring weights for partially conditional rate-based analyses \citep{Datta2002, Cook2009}. 
This enables consistent estimation of the $q_{kl}(t|X)$ in (\ref{qkl}) governing the process
in Figure \ref{fig-intensity-based-models}(a) and from this, weighted Aalen-Johansen estimates of state occupancy probabilities $P(Z(t)=k|X)$ can be obtained as described by \citet[Section 3.4.2.]{Cook2018}. This allows estimation of estimands such as \textit{i)}- \textit{ii)} described above.

We note that an additional more subtle requirement for conditionally independent loss to followup  given in (4b) of \cite{Lawless2019} is expressed here in terms of Figure \ref{fig-intensity-based-models}(c) as
\begin{equation}
\label{ind2}
\lambda_{k' l'}(t | H^\circ(t)) = \lambda_{kl}(t | H(t)) \; ,
\end{equation}
\noindent
for $(k, l) = \{(0,1), (0,2), (1,2) \}$.
This assumption implies that among those recruited and randomized to treatment,  the intensities of the illness-death process are the same for those under
followup as those who have been lost to followup.
This assumption cannot be verified in the absence of data following loss to followup.
Such data can sometimes be obtained through tracing studies \citep{Lawless2019}, but these are seldom done in clinical trials.  Moreover,  subjects in a trial receive care and treatment by protocols which do not apply after withdrawal and so one would not expect disease dynamics to remain the same as if the person had remained in the trial.  Thus, the main objective should be to protect against dependent loss to followup through violation of (\ref{ind1}) by using weights as described above.

We stress that when some degree of premature LTF is a feature of a trial, estimands such as \textit{i)}- \textit{ii)} above must address LTF. In particular, consider the commonly used event times $T,T_1,T_2$ associated with Figure \ref{fig-intensity-based-models}(a); here $T$ is the exit time of state $0$ and $T_1,T_2$ are the times of entry to states $1$ and $2$ respectively. Interpretation of $P(T > t|X)=P(Z(t)=0|X)$ is clear in the setting of Figure \ref{fig-intensity-based-models}(a), but when the observable process is that in Figure \ref{fig-intensity-based-models}(b),  $P(T > t|X) = P(Z^{\circ}(t) = 0|X)$ represents the probability that an individual is event-free, alive, and not lost to followup at time $t$.
If LTF arises because of treatment discontinuation due to lack of efficacy or adverse effects, this represents a reasonable strategy.
If LTF is driven by fixed or time-varying biomarkers associated with event occurrence, then LTF is dependent and incorporating LTF into a composite event
(implicit in modeling the sojourn time in state 0 of Figure \ref{fig-intensity-based-models}(b)) is one solution.
If there is no evidence that LTF is marker-dependent then LTF can simply be treated as independent right censoring and analyses can be based on
Figure \ref{fig-intensity-based-models}(a).
Similar issues arise when considering $P(T_1 \le t|X)$ and $P(T_2 \le t|X)$, which in Figure \ref{fig-intensity-based-models}(b) represent
the probability that a subject has entered states 1 or 2 respectively prior to time $t$, while under followup.

\subsubsection{Treatment discontinuation, no loss to followup } \label{sec4.2.2}
In some settings the intercurrent event may be a toxicity-related event leading to discontinuation of the assigned treatment, but with the subject remaining under followup, perhaps with a change in treatment.
This constitutes a type 2 IE with observed data as in Figure \ref{fig-intensity-based-models}(c).
In this case, as in Section \ref{sec4.2.1}, we should incorporate this expanded process when defining features and estimands of interest.  
For event-free survival time $T$, for example, we should consider $P(T>t)=P(Z^\circ(t) \in (0,0'))$.  For treatment group comparisons that relate to 
observable processes, we support intention-to-treat comparisons \citep{montori2001},  which here might for example be based on $\beta(\tau) =
 P(T>\tau|X=1) - P(T>\tau|X=0)$, with $X$ representing treatment assigned at randomization.  This in our opinion gives a more relevant estimand concerning treatment efficacy in most settings than another option that has been proposed,  which is to artificially censor subjects when they cease the initial treatment;  in that case, Figure \ref{fig-intensity-based-models}(b) would be used.

In many settings decisions about treatment switches are related to biomarkers which are used in monitoring subjects and in this case, expanded models may include a marker process. As an illustration, we consider a time-dependent marker $\{W(t), 0< t\}$ where $W(t)$ reflects some aspect of disease severity,  for example   a marker of bone formation or destruction (e.g. bone alkaline phosphotase) in cancer patients with skeletal metastases.
We let $H(t) = \{Z(s),  0 < s< t\}$  be the history of the illness-death process in Figure \ref{fig-intensity-based-models}(a) and
${\cal W}(t) = \{W(s),  0 < s< t\}$  be the history of the marker process alone.
The expanded history ${\cal H}(t) = \{Z(s), W(s), 0 < s< t, X\}$  includes the multistate and marker process histories along with the treatment covariate.
We let  $\lambda_{kl}(t|{\cal H}(t))$ denote the transition intensities for $(k,l)$ in ${\cal S}$ for the illness-death process of
Figure \ref{fig-intensity-based-models}(a), now accommodating dependence on the marker process.
Likewise we let $\lambda^\circ_{kl}(t|{\cal H}^\circ(t))$ denote the intensities for transitions among the pairs of illness-death states of
Figure \ref{fig-intensity-based-models}(c) for $(k,l) \in \{(0,1), (0, 2), (1, 2), (0',1'), (0', 2'), (1', 2')\}$ where ${\cal H}^\circ(t) = \{Z^\circ(s), W(s), 0 < s< t, X\}$.
Modeling the intensities $\lambda_{jj'}(t|{\cal H}^\circ(t))$  for $j \rightarrow j'$ transitions corresponding to the occurrence of the IE can offer important insights concerning the types of individuals who cannot tolerate the study medication.  A major difficulty in this situation, however, is the need to model the marker process in order to define and calculate marginal process features such as $P(T > t|X)$ that can be used to define estimands.  A discussion of this area is beyond our present scope; \cite{Cook2022} provide an illustration.

\subsubsection{Introduction of rescue treatment, no loss to followup} \label{sec4.2.3}
In cancer clinical trials it is common for individuals to receive rescue therapy, often after evidence of disease progression, with followup continuing after the 
rescue therapy has been introduced.
The decision to prescribe rescue therapy will often be guided by marker processes, perhaps in combination with information on the illness-death process.
In the case where rescue therapy is only  introduced upon disease progression (entry to state 1) the $0 \rightarrow 0'$ transition intensity in 
Figure \ref{fig-intensity-based-models}(c) is zero.
In such a setting estimation of event-free survival probabilities or cumulative incidence of disease progression is unaffected but overall survival probabilities 
are affected.
Modeling the $0 \rightarrow 0'$ and $1 \rightarrow 1'$ intensity functions will provide insight into the kinds of individuals who are prescribed rescue therapy.
As in Section \ref{sec4.2.2} the intention-to-treat principle considers the full process in
Figure \ref{fig-intensity-based-models}(c) and is preferred for treatment comparison.

\begin{figure}[!ht]
\noindent
\begin{minipage}[t]{0.485\textwidth}
\begin{center}
\includegraphics[scale=0.43]{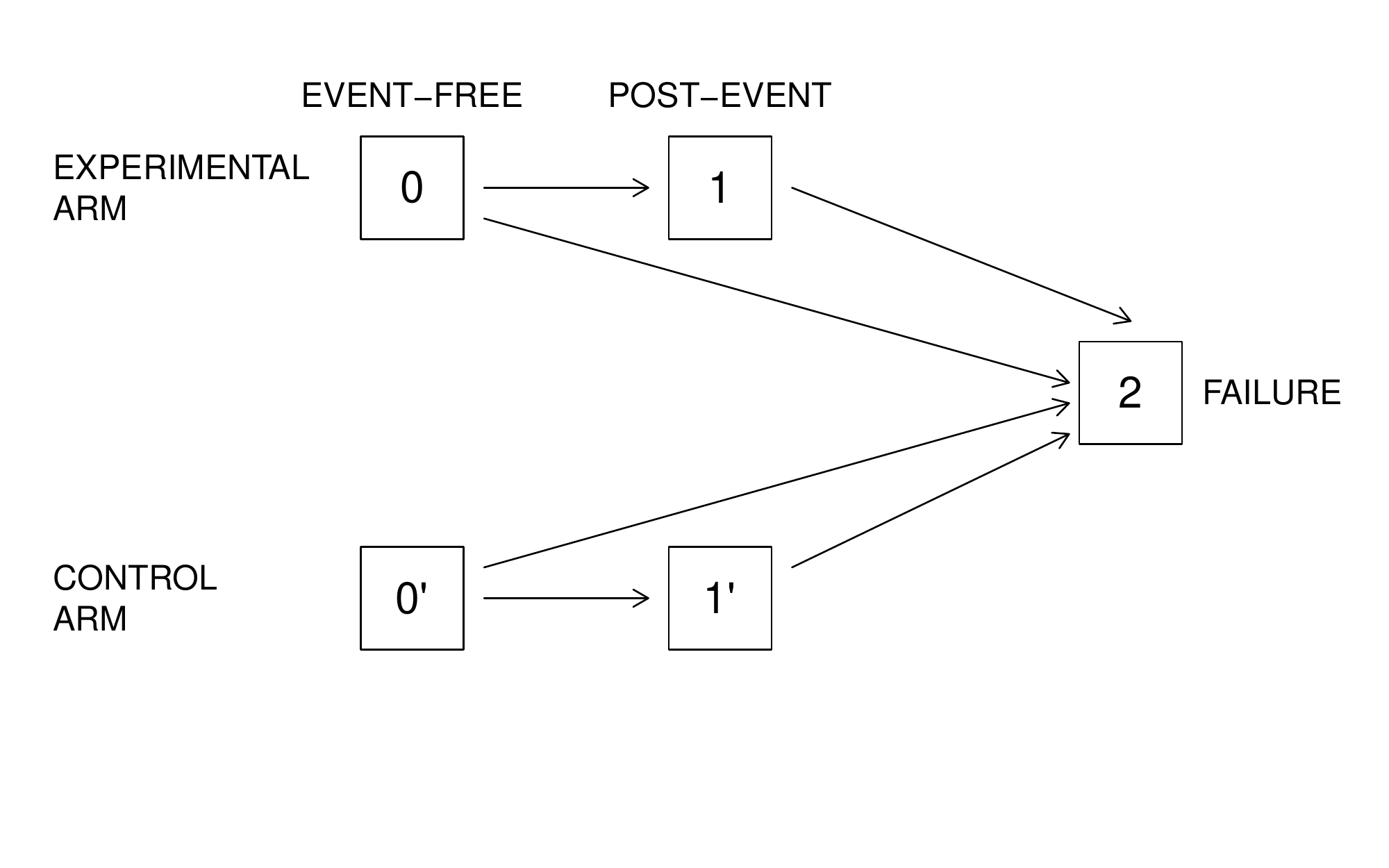} \\
(a) Illness-death processes for two treatment arms
\end{center}
\end{minipage}
\hfill
\noindent
\begin{minipage}[t]{0.485\textwidth}
\begin{center}
\includegraphics[scale=0.43]{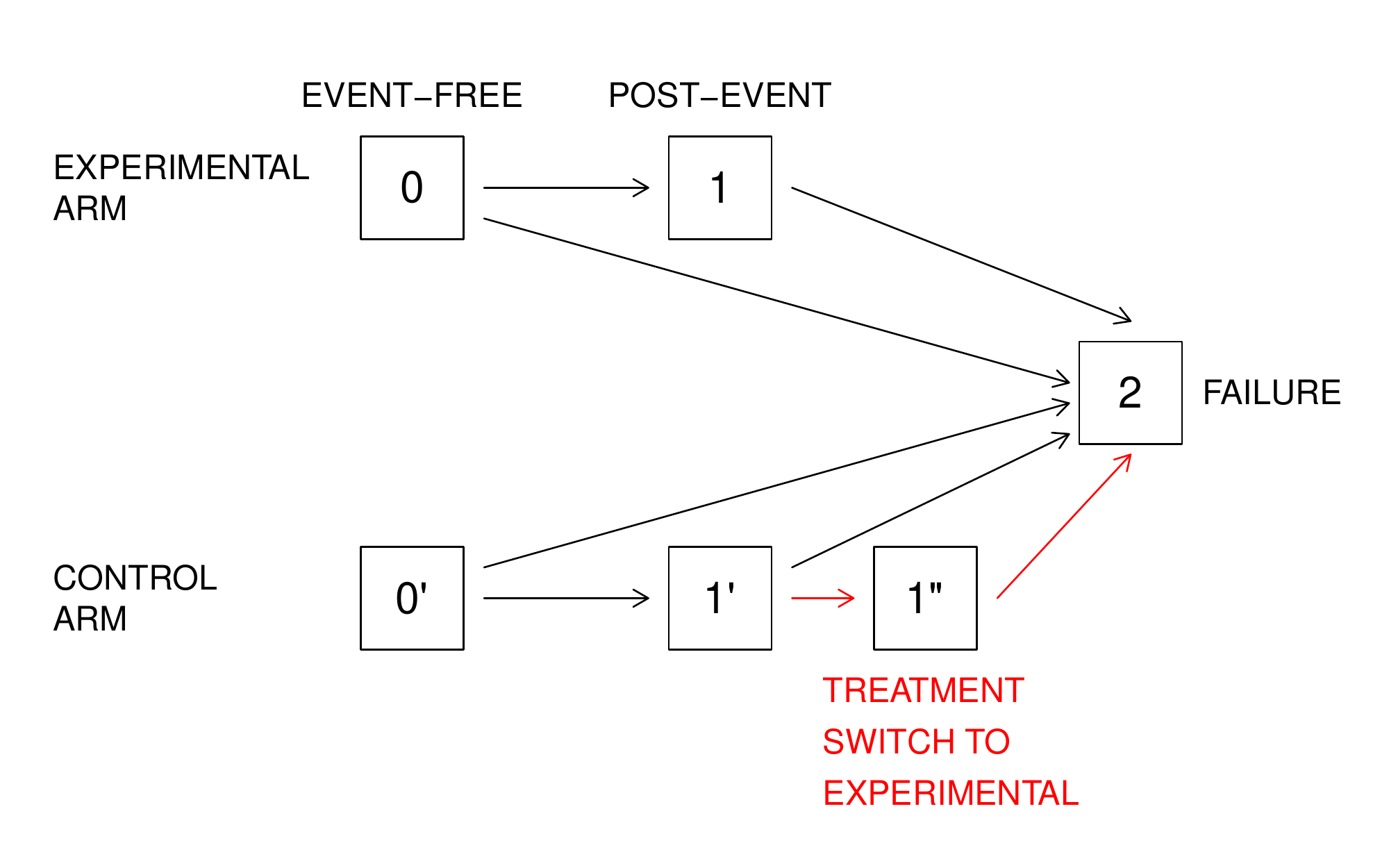} \\
(b) Illness-death processes with possible control to experimental arm crossover
\end{center}
\end{minipage}
\caption{Multistate models for a process of interest (panel (a))
and an expanded process incorporating a post-randomization intervention
for the control arm (panel (b)).}
\label{fig-marker-trt2}
\end{figure}

In some trials individuals may be switched from their assigned treatment to the treatment of the other arm.  
Most often individuals in the control arm are prescribed the experimental treatment; in cancer trials this is often done following cancer progression \citep{watkins2013}. 
Figure \ref{fig-marker-trt2}(a) is a multistate diagram for an illness-death process $\{Z(t), 0 < t\}$ with separate states for two treatment groups, with overall survival the 
feature of interest.
\cite{henshall2016}, \cite{latimer2019b} and \cite{latimer2019} for example consider this setting, though not with our expanded model.
Our preference is again to formulate intensity-based models of observable event times and make clear and explicit assumptions about the disease and treatment process.
In Figure \ref{fig-marker-trt2}(b) we add a state $1''$ which can be entered from state $1'$ (indicating progression under the control therapy)
to reflect the introduction of the experimental treatment, resulting in the six-state multistate process $\{Z^{\circ}(t), 0 < t \}$. 
As in Sections \ref{sec4.2.2} and \ref{sec4.2.3} the intensity functions for the $1' \rightarrow 1''$ transiton 
give insights into the kinds of individuals switching from the control to the treatment arm.
In this setting we again favour use of the intention-to-treat principle in conjunction with nonparametric estimation of $P(Z^{\circ}(t)=2|X)$.

We note that it is common to see analysis of individual or composite event times such as $T,T_1$ and $T_2$ based on Cox proportional hazards models, with the 
estimand $\beta$ defined as the log hazard ratio for treament to control groups.  
Even if we interpret $\beta$ causally through an analogous transformation model as discussed earlier,  this should be accompanied by an assessment of the 
model's adequacy.  
With either a model-based or nonparametric estimand as in \textit{i)}- \textit{ii)}, secondary intensity-based analysis is crucial to an understanding of factors 
producing an observed marginal effect.
Another type of secondary analysis is to examine conditional probabilities such as
$P(Z^\circ(t)=0| Z^\circ(t) \in (0,1,2),X)$ or $P(Z^\circ(t)=2| Z^\circ(t) \in (0,1,2),X)$; 
these summarize event occurrence up to time $t$ for persons not experiencing the IE by that time. 
Such probabilities can be estimated nonparametrically and although they are not suitable for causal interpretation because they condition on events that 
may be dependent on treatment or unobserved confounders, they and corresponding estimates of $P(Z^\circ(t) \in (0,1,2)|X)$ provide useful summary information.
There has been discussion in the literature and in the ICH E9 (R1) guidance document \citep{iche9-2017} about the use of principle stratification and estimation of the so-called 
survivor average causal effect.  
These methods condition on similar events but use counterfactuals, so do not meet our objective of real world interpretation; Section \ref{sec5.1} discusses this further.

\subsubsection{Surgical prevention of stroke-related events in the NASCET study}
\label{secnascet}
\begin{figure}[!ht]
\begin{center}
\begin{minipage}[t]{0.45\linewidth}
\includegraphics[scale=0.4]{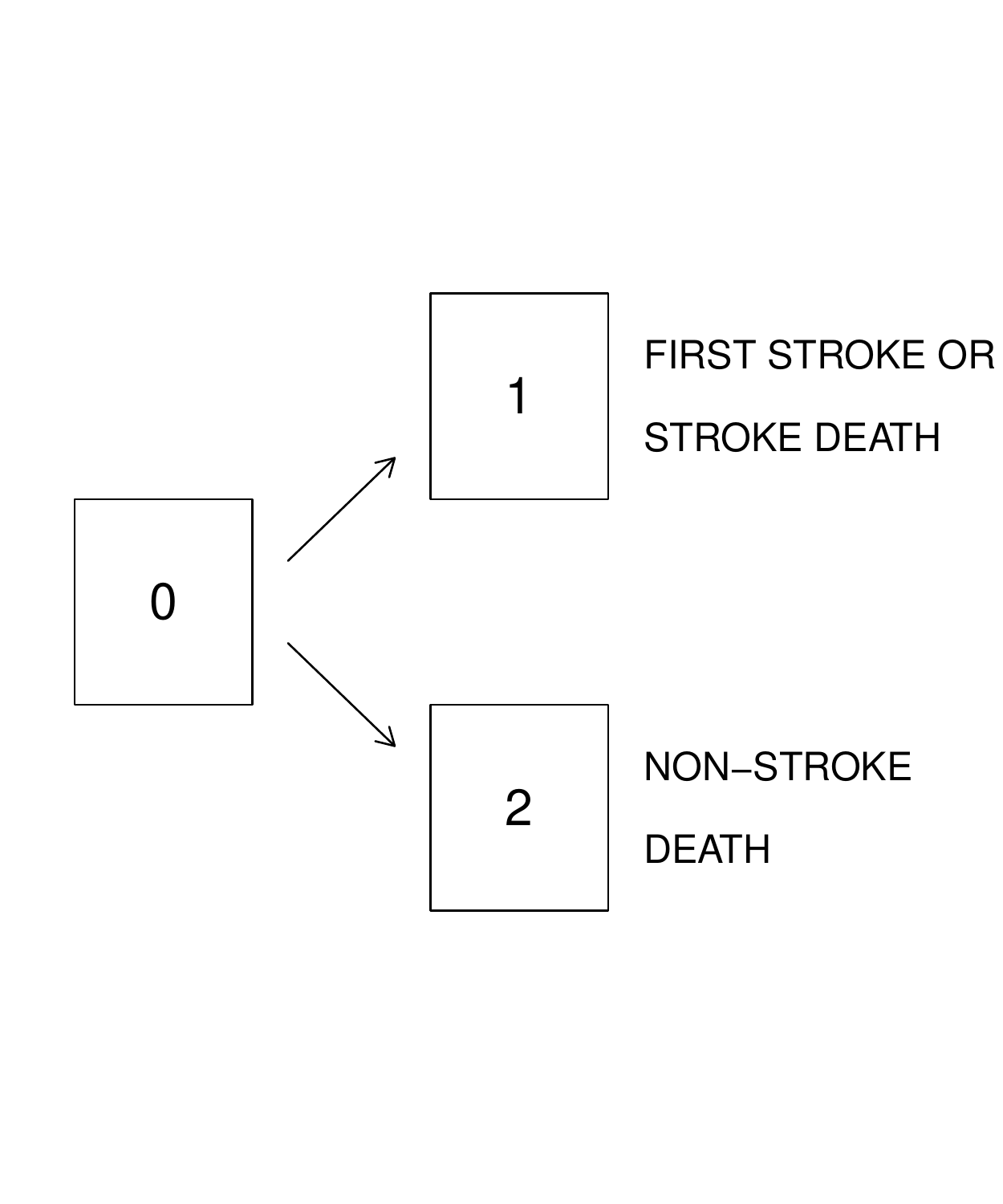} \\
{\footnotesize (a) A simple competing risk model for an intention-to-treat analysis}
\end{minipage}
\begin{minipage}[t]{0.5\linewidth}
\includegraphics[scale=0.4]{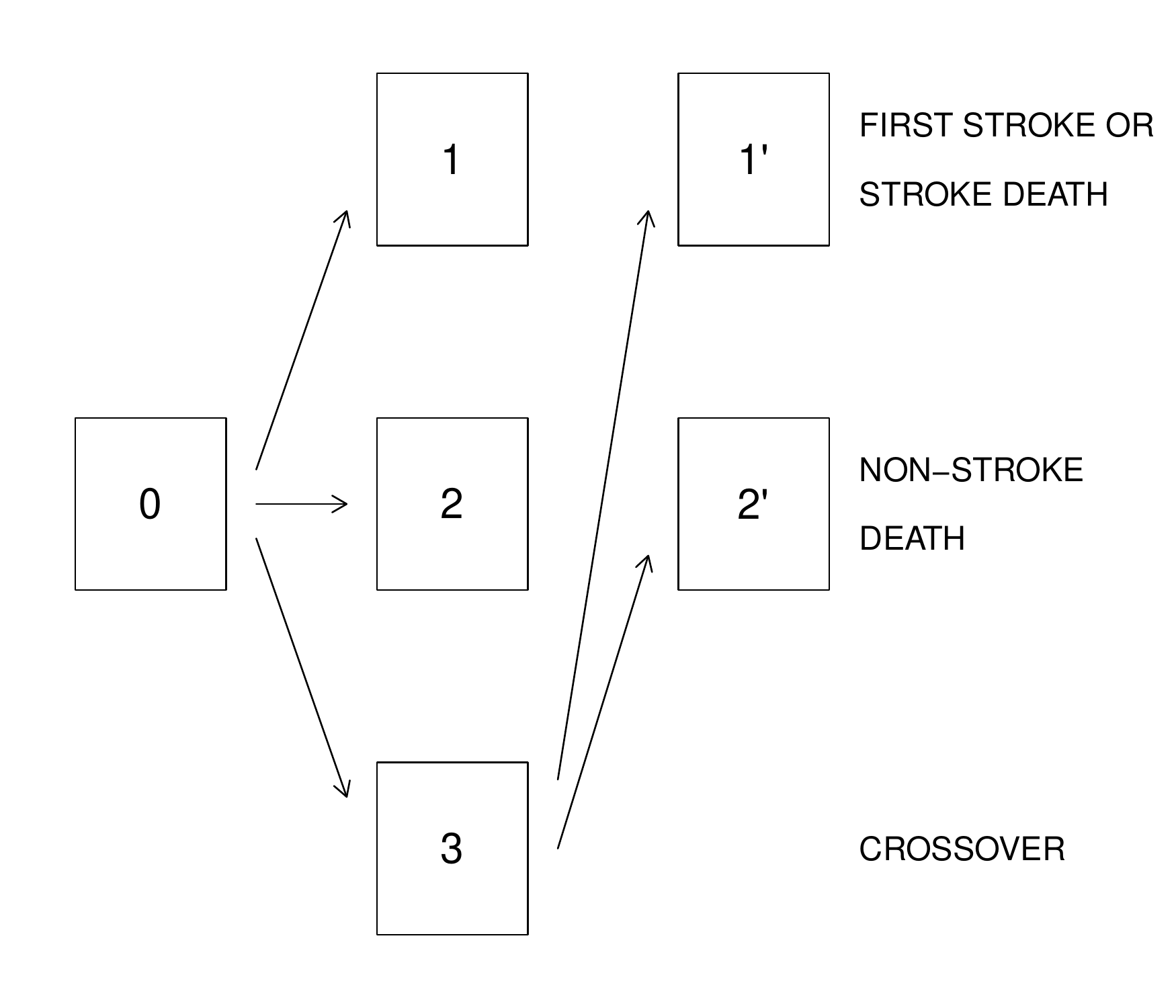} \\
{\footnotesize (b) A multistate model accommodating information on medical-surgical crossover}
\end{minipage}
\caption{Multistate diagrams for analyses of stroke or stroke-related death and non-stroke related death.}
\label{fig-nascet-ms-death}
\end{center}
\end{figure}

Here we consider data from the NASCET trial \citep{Barnett1998,NASCET-STROKE1991}
in which we focus on individuals randomized to medical care $(n=1118)$ or carotid endarterectomy $(n=1087)$ in the stratum with moderate ($< 70$\%) stenosis.
We consider for illustration an analysis involving the response of stroke-related events (i.e. stroke or stroke-related death).
Figure \ref{fig-nascet-ms-death}(a) shows the multistate diagram for a competing risks process $\{Z(s),  0 < s\}$ where 
state 0 represents the condition of being stroke-free and alive,
state 1 is entered up the occurrence of a stroke or stroke-death, and 
state 2 is entered upon a non-stroke death.
Note that entry to state 1 due to a non-fatal stroke may be followed by a stroke death or non-stroke death but we consider this simplified model, which is 
 relevant for an analysis of the composite event of non-fatal stroke or stroke death; then the only competing event is non-stroke death.

\begin{figure}[!ht]
\begin{center}
\includegraphics[scale=0.55]{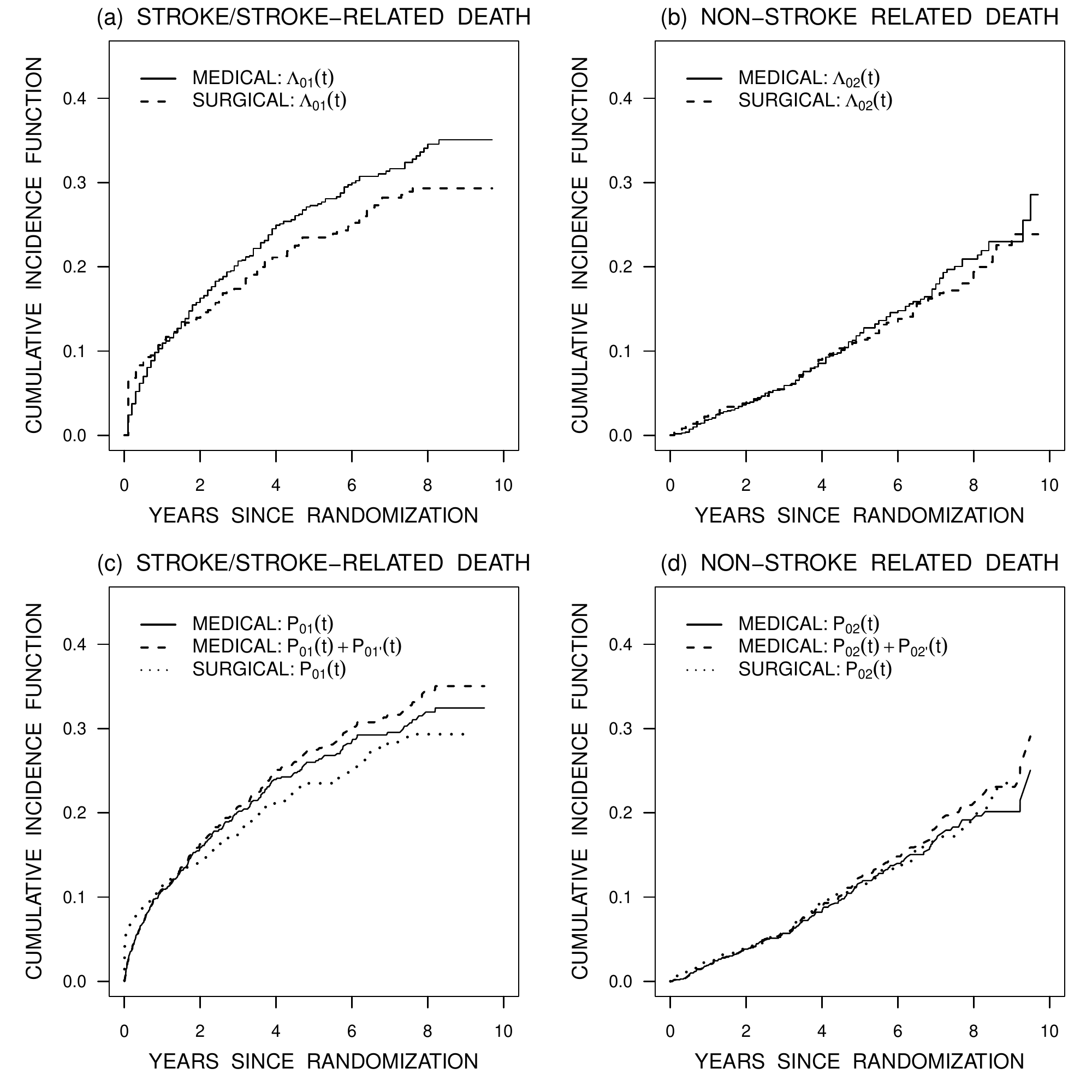}
\caption{Estimated cumulative incidence functions for stroke/stroke-death and non-stroke  death in Figure \ref{fig-nascet-ms-death}(a) (top row) and
Figure \ref{fig-nascet-ms-death}(b) (bottom row); Aalen-Johansen estimates are used for the state occupancy probabilities in panels (c) and (d).}
\label{fig-nascet-death-cif}
\end{center}
\end{figure}

As noted in Section \ref{nascet-intro}, a number of individuals randomized to receive best available medical care crossed over to undergo carotid endarterectomy.
The reasons recorded for this included non-fatal stroke, further narrowing of the carotid artery based on angiograph examination,
both stroke and angiographic progression, and other reasons\footnote{unfortunately while this information was recorded at the time 
of the trial this data are not available.}.
Figure \ref{fig-nascet-ms-death}(b) shows the more complex multistate process relevant for individuals in the medical arm
which accommodates the crossover to carotid endarteractomy.
Under the intention-to-treat framework the cumulative incidence functions for the model in Figure \ref{fig-nascet-ms-death}(a) 
can be estimated for both the surgical and medical arms; 
these are given in 
Figure \ref{fig-nascet-death-cif}(a) 
for the composite event of stroke or stroke-related death and 
Figure \ref{fig-nascet-death-cif}(b) for non-stroke death.
Carotid endarterectomy can dislodge plaque into the circulatory system and cause stroke, so there is a perioperative period of high risk of stroke reflected 
by the steep increase in the cumulative incidence function estimate in Figure \ref{fig-nascet-death-cif}(a) for the surgical arm. 
Following this perioperative period the cumulative risk of stroke or stroke-death is lower in the surgical arm, so that ultimately there is
a 6-8\% absolute reduction in the risk of the composite event in the surgical arm by 8 years.
Figure \ref{fig-nascet-death-cif}(b) shows very similar cumulative incidence functions for non-stroke death in the two arms.

The intention-to-treat principle is normally justified on the basis that any interventions following randomization are part of routine care.
To help in understanding and communicating the relevance of trial findings, it is necessary to clearly describe what constitutes routine care.
For the NASCET study involving individuals deemed at moderate risk, the situation is more complicated.
We focus here on the moderate risk stratum of NASCET which was treated as a separate, parallel study to one involving 
individuals designated as high risk due to having greater than 70\% stenosis  of the carotid artery at the time of randomization.
The trial involving high risk patients started at the same time as the trial of moderate risk patients, but was stopped early when strong evidence 
emerged at an interim analysis of a benefit of carotid endarterectomy -- this occurred during the conduct of the trial for the moderate risk patients.
A second point is that individuals at moderate risk at the time of randomization to medical care may have experienced a progression of their carotid stenosis 
over the course of followup to the point that it exceeded 70\% -- this then qualified them as high risk patient. 
Following the publication of trial results for high risk patients, the standard of care changed to include carotid endarterectomy. 
As a result some patients randomized to medical care in the trial of moderate risk individuals may have experienced stroke when carotid endarterectomy 
was not part of standard of care, whereas
others may have been randomized, progressed to qualify as high risk, and received carotid endarterectomy while stroke-free. 
The risk of stroke following surgery in this latter group of would be influenced by the surgical procedure.
The results of an intention-to-treat analysis are difficult to interpret when the standard of care changes during the course of a study -- a 
more detailed analysis is warranted to understand risks and associated effects.

Figure \ref{fig-nascet-ms-death}(b) shows the multistate diagram for a more detailed characterization of the course following randomization.
States 1 and 2 again correspond to stroke or stroke-death (state 1) and non-stroke death (state 2) but state 3 is entered upon medical-surgical crossover for 
individuals in the medical arm.
By orienting the states in this competing risk arrangement we note that state 1 is now entered only if the stroke or stroke-death occurs prior to surgical crossover,
and likewise for the non-stroke death.
Medical-surgical crossover can be followed by stroke/stroke-death or non-stroke death so states $1'$ and $2'$ represent these events.
Letting $\{Z^{\circ}(s), 0< s\}$ denote the multistate process in Figure \ref{fig-nascet-ms-death}(b), we note that the Aalen-Johansen estimate of the 
transition probability matrix facilitates a decomposition of the cumulative incidence function estimates in Figures \ref{fig-nascet-death-cif}(a) and (b). 
If $X=1$ for an individual randomized to receive surgery and $X=0$ if they are randomized to medical care, then we note that $P(Z^{\circ}(t)=1|X=0) + P(Z^{\circ}(t)=1'|X=0)$ 
is the cumulative incidence function for stroke/stroke-death estimated in Figure \ref{fig-nascet-death-cif}(a).
The separate Aalen-Johansen estimates of $P(Z^{\circ}(t)=1|X=0)$ and $P(Z^{\circ}(t)=1'|X=0)$ in Figure \ref{fig-nascet-death-cif}(c) show that the excess risk of 
stroke/stroke-death in the medical arm evident in Figure \ref{fig-nascet-death-cif}(a)
is made up of risk for those not crossing over, and risk from those crossing over to receive surgery.
We cannot attribute this added risk to the surgery itself due to confounding by indication; we may expect an elevation in risk shortly after surgery and a reduction
in risk following a perioperative period, but a critical point is that those crossing over to receive surgery are selected in a dynamic way in response to their disease
course. 

\begin{figure}[!ht]
\begin{center}
\includegraphics[scale=0.55]{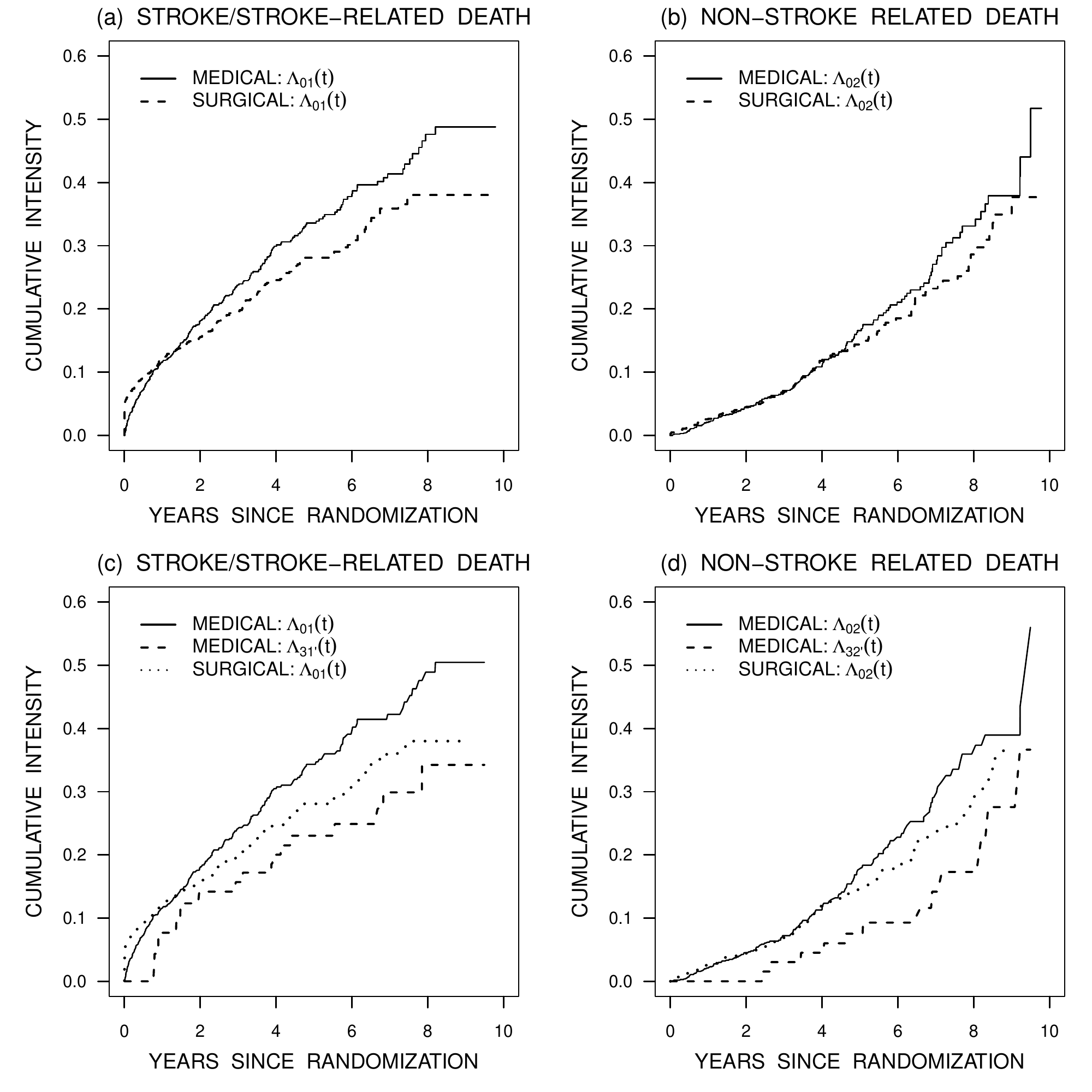}
\caption{Nelson-Aalen estimates of the cumulative intensities for 
selected transitions based on Figure \ref{fig-nascet-ms-death}(a) (top row) and 
Figure \ref{fig-nascet-ms-death}(b) (bottom row).}
\label{fig-nascet-death-cumint}
\end{center}
\end{figure}

To explore this more fully we plot Nelson-Aalen estimates of the cumulative transition intensities in Figure \ref{fig-nascet-death-cumint} for the multistate models of 
Figure \ref{fig-nascet-ms-death}.
The Nelson-Aalen estimates in Figures \ref{fig-nascet-death-cumint}(a) and (b) correspond to those of Figure \ref{fig-nascet-ms-death}(a) and
we again see evidence of the perioperative period of elevated risk following surgery, with similar cumulative intensities for non-stroke death.
The Nelson-Aalen estimates of the cumulative transition intensities in Figure \ref{fig-nascet-death-cumint}(c) 
are for transitions in Figure \ref{fig-nascet-ms-death}(b).
Interestingly, the slope of the estimated
$0 \rightarrow 1$ cumulative intensity for the surgical arm and the 
$3 \rightarrow 1'$ cumulative intensity for those in the medical arm following surgical-crossover are very similar; 
the steepest cumulative intensity estimate is for the medical patients who did not crossover.
A similar decomposition is given in Figure 
\ref{fig-nascet-death-cumint}(d) for the endpoint of non-stroke death. 

Intensity-based treatment comparisons could be based on regression modeling but the crossing cumulative  intensities seen in Figure \ref{fig-nascet-death-cumint}(a) 
mean that proportional cause-specific hazards models will not be suitable. 
A detailed analysis would make use of a model allowing crossover between treatment arms and this would ideally   incorporate information on the disease course including
blood pressure, cholesterol measurements and angiographic assessment of carotid stenosis. 
Data on such variables are unfortunately not available so we do not explore this, but note that there 
is much ongoing work on  the assessment of randomized treatments when hazards or cumulative incidence functions cross, and when 
treatment switching occurs.

\subsection{Estimands for recurrent and terminal events based on marginal rate functions} \label{sec4.3}

Other disease processes involve non-fatal events that may occur repeatedly. 
In cardiovascular research, for example,  events such as myocardial infarction (MI),  stroke and admission to hospital due to heart failure may each recur \citep{Schmidli2021, Toenges2021,Furberg2021}. 
Competing terminal events such as death or loss to followup are also common.  
The risk of recurrent skeletal complications in the bone metastases trial in Section \ref{bonemets-study} is, for instance, terminated by death; see Figure \ref{sre}.  
Section 6.4 of \cite{Cook2007} and \cite{Cook2009} give a thorough treatment of such processes involving recurrent and terminal events; 
for a recent review of methods in the context of randomized trials see \cite{Furberg2021}, \cite{Toenges2021} and \cite{Mao2021b}.  
In such settings marginal estimands based on cumulative counts for recurrent non-fatal events, or counts comprised of either fatal or non-fatal events are widely used. In connection with the LEADER (Liraglutide Effect and Action in Diabetes: Evaluation of Cardiovascular Outcome Results) trial,  
\cite{Furberg2021} consider a recurrent  composite event with components consisting of non-fatal stroke, non-fatal MI or cardiovascular (CV) death.  
Such a composite recurrent event is motivated by a desire to synthesize information about treatment effects across multiple types of events, but 
as with any composite endpoint clear interpretation requires modeling treatment effects on the component event types. 

{
If primary interest lies in the effect of treatment on non-fatal recurrent events, the proportional means model of \cite{Ghosh2002} can be used. 
It is based on the marginal rate function for the non-fatal event,  recognizing that further non-fatal events 
cannot occur after a terminal event; it is analogous to a cumulative incidence function in this way, but accommodates recurrent events.
The ratio of rate or mean functions  satisfies the principles for estimands given in Section \ref{sec2.1} and has a descriptive causal interpretation. 
\cite{Mao2016} proposed a multiplicative model based on the rate function and corresponding mean function for the composite counting process for fatal and non-fatal events combined. 
Estimands based on the ratio of such mean or rate functions also satisfy our principles of Section \ref{sec2.1} and have a descriptive causal interpretation. 
By including  terminal events in the composite counting process, the Mao-Lin approach may be suited to settings where treatment is needed to avoid both serious non-fatal events and  terminal events. 
There are however the usual caveats: 
(a) intensity-based  models should be fitted in secondary analyses to gain understanding of the mechanisms
leading to any observed treatment effect; Cox models and other regression models for terminal event intensities are useful in understanding connections between non-fatal and fatal events, and 
(b) the proportionality of rate or mean functions inherent to the Ghosh-Lin and Mao-Lin models should be checked. In applications and ongoing numerical investigations we have found that both the Ghosh-Lin and Mao-Lin models provide a reasonable summary of
treatment effects in a range of settings \citep{buhler22}.
}

\section{Additional remarks} \label{sec5}

\subsection{Further remarks on potential outcomes} \label{sec5.1}

{
Potential outcomes have played a central role in the development of causal inference theory and methods over the last several 
decades with their origins in cross-sectional or simple longitudinal settings \citep{rubin2005, hernan2020}.
The potential outcome framework 
conceptualizes the existence of outcomes for each treatment under consideration for each individual.
In a two treatment trial this leads to a pair of potential outcomes which could, conceptually, be compared for each individual in the trial.
In trials where each individual receives just one treatment, the pairs of 
potential outcomes and "causal" measures such as their difference are thus not observable in the real world.
The potential outcome corresponding to the treatment received is revealed, while the other is counterfactual.
Some find potential outcomes useful for conceptualizing independence conditions, and this is reasonable in settings where one could, 
both in theory and in practice, randomize individuals to one of two treatments under study \citep{hernan2020}. 
We note that thinking in terms of potential outcomes is increasingly common in connection with clinical practice, but in our view this does not line up in most settings with clinicians' views. For example, 
\cite{bornkamp2021} say 
``Assume a treating physician is deciding on a treatment to prescribe. 
Ideally she would make that decision based on knowledge on what the outcome for the patient would be if given the control treatment, $\ldots$, and what the outcome 
would be under test treatments$\ldots$''. The term "ideally" notwithstanding, our view is that physicians think and counsel patients not in terms of fixed potential outcomes under alternative
treatment options but in terms of probabilities that are 
based on data from trials and observational studies. For example, a physician might inform a breast cancer patient following surgery that the (estimated) probability of a recurrence during the next five years is $0.10$ with no adjuvant hormone therapy and $0.08$ with adjuvant therapy.

We believe that the counterfactual framework can also lead to specification of target estimands of dubious scientific relevance when used in conjunction with the concept of principal strata.
Consider the problem of assessing the effect of a new intervention versus standard care on an outcome that can only be measured in individuals 
who are alive; an example is the measurement of neonatal outcomes among infants born in studies of different techniques for in vitro 
fertilization \citep{snowden2020}. 
\cite{rubin2006} proposed estimands based on women who would experience a live birth under either of two treatments under study; this is referred to as the survivor average causal effect, with those surviving under either treatment comprising one of the four so-called "principal strata".
Aside from concerns about counterfactual outcomes not being observable in the real world,  such subgroups are not identifiable (observable) from the available data. Such issues have been discussed by others. 
As noted by \cite{lipkovich2022} for example, ``a major challenge of using principal stratification $\ldots$ is its counterfactual nature 
that requires strong assumptions to be able to identify and estimate treatment effect within principal strata''.
Moreover, as noted by \cite{scharfstein2019}, the survivor average causal effect does not convey the effect of treatment on 
all of those randomized, and the subset of the population in the principle stratum of interest may be small.
Issues of competing risks also arise in settings with intercurrent events \citep{iche9-2017}
such as early withdrawal,
introduction of rescue medication,
crossover to the complementary arm, and death.
\cite{hernan2018} provide a thoughtful discussion of the ICH Addendum and cautionary remarks about potential outcomes and principal strata.
Finally, \cite{Andersen2012.1} caution against conditioning on the future in analysis of life history processes if interest 
lies maintaining clear real world interpretation of estimands; the survivor average causal effect violates this tenet.
Preferable strategies for dealing with intercurrent events include ignoring their occurrence in an intention-to-treat analysis 
targetting the effect of a "treatment policy" (i.e. the effect of prescribing one treatment versus the other at the time of study entry).
The intercurrent event can also be incorporated into a composite endpoint, which can make sense if the event of interest and the intercurrent event
are both undesirable.

Disease processes and patient management are by nature dynamic, and 
time plays a central role. \cite{aalen2012-2} among others have highlighted the importance of time, and of modeling data in ways that recognize 
the evolutionary nature of disease processes. This way of thinking is aligned with the principle of "staying in the real world" but, as we have discussed,  complicates the identification of relevant marginal features which can deliver causal inferences based on randomization to treatment.  
The dynamic nature of  processes also makes  the notion of pre-existing counterfactual outcomes artificial; probabilistic models are needed for real world processes.
}

\subsection{The utility of utilities in defining causal estimands} \label{sec5.2}

In pharmaceutical research a considerable amount of discussion typically takes place between companies and regulators regarding study protocols, with much of it related to the choice of primary endpoint, specification of the estimand, and the associated analyses. As we have discussed, it can be challenging to define estimands when individuals in a clinical trial are at risk of different types of clinically important  events, some of which may be recurrent.
Evidence must be synthesized  to make a decision about the relative value of an experimental treatment, but
initial discussions usually aim to set a protocol and analysis plan wherein the criteria for trial success and the path to regulatory approval for a treatment is clear.  Trial planning is facilitated by considering a marginal estimand for which observed treatment effects can be given a causal interpretation.
We have stressed that secondary, possibly intensity-based analyses provide  a fuller understanding of the response to treatment. Conclusions about a treatment, and its approval,  should be based on  the totality of evidence based on the primary analysis and important secondary analyses.

Various approaches for combining information across multiple event types have been proposed.  Compound events composed of two or more types are common; this was discussed in Section \ref{sec4.3} for the case of recurrent non-fatal and fatal events. A recently proposed approach with complex processes is the win ratio \citep{pocock2012}. 
It involves a rule by which two process outcomes (or "paths") can be compared, with either one path being adjudged superior (the "winner") or them being declared tied.  
With such a rule we can consider all possible pairs $(i,j)$ of subjects, where $i$ is from one treatment group and $j$ from the other, and declare either a winner or a tie.
We can then obtain the proportion of (non-tied) pairs where the winner was in the experimental arm; the win ratio is  the ratio of this proportion and its complement.
The rule for ranking  process outcomes or paths depends on the context.  In an illness-death process, for example,  if one subject dies after the other, they are the winner; if neither subject dies then the one who enters the non-fatal event (illness) state later is the winner.  Subjects who are both in the healthy state at administrative end of followup are treated as tied.  We will return briefly to this setting at the end of this section. 
Stratified versions of the win ratio and other extensions have also been  
developed \citep{luo2015, oakes2016, mao2019, mao2021}.

An alternative approach in complex disease processes is to use utilities.  This involves defining a numerical utility score associated with events or with time spent in specific states, and then computing the total utility score for each subject across the duration of the trial.  This approach has the advantage of transparency and it produces marginal features and estimands that admit descriptive causal interpretations.  A disadvantage is the need to specify utility scores; there may be a lack of consensus.
We note that many other approaches implicitly involve utilies or assumptions about relative utility score.  This is true of win ratios as well as composite endpoints. 
In a composite endpoint, for example, the component events are implicitly of equal importance if a failure event is deemed to have taken place when any one of the component 
events occurs.

\begin{figure}[!ht] \begin{center}
\includegraphics[scale=0.50]{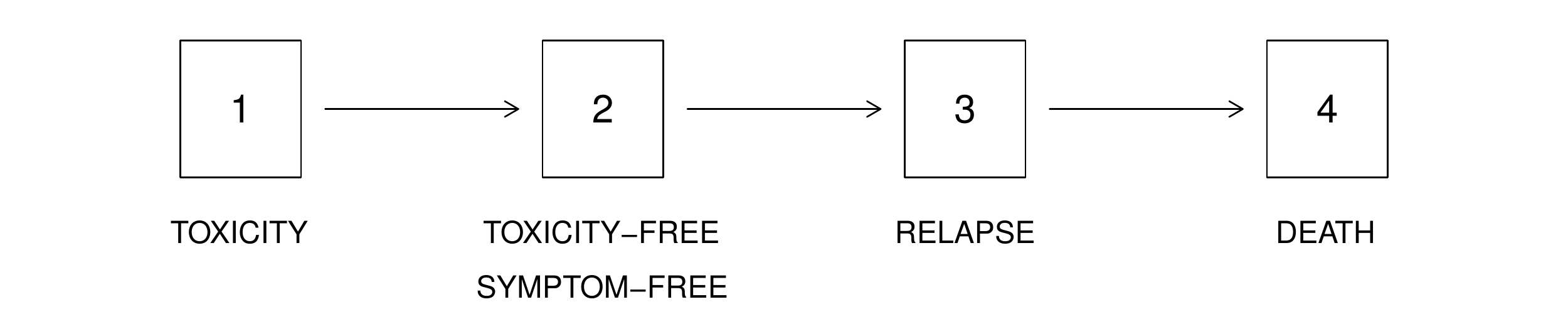}
\caption{A four-state multistate model distinguishing the phases of toxicity from chemotherapy, the time without toxicity or symptoms, relapse and death, for computation
of the Q-TWiST measure of \cite{Gelber1989}.} \label{QoL}
\end{center} \end{figure}

Utilities have frequently been used to address quality of life outcomes.  An example was in a randomized clinical trial of adjuvant chemotherapy for breast cancer patients conducted by the 
International Breast Cancer Study Group (IBCSG) \citep{LBCSG1988}.  
A total of 1,229 patients were randomized to receive either short duration ($n=413$) or long duration ($n=816$) chemotherapy and followed for a median of about 
seven years.
Patients experience toxicity during chemotherapy but intervals during which they are toxicity-free and relapse-free typically follow the completion of chemotherapy;
patients may of course relapse and die following such an interval. 
To assess the impact of short versus long duration chemotherapy,  overall survival analyses can be conducted. 
The therapeutic issue under investigation here is, however, whether the greater anti-tumour effect of longer duration chemotherapy outweighs the protracted 
periods of toxicity-related symptoms and poorer quality of life. To explore the impact of treatment on the sojourn times in the different stages of response, \cite{Gelber1989} proposed partitioning the survival time into different stages according to a four-state progressive model where
state 1 is occupied during the period of toxicity due to chemotherapy,
state 2 is occupied following chemotherapy when individuals are toxicity-free and symptom-free,  state 3 is entered when symptoms reemerge upon relapse, and 
state 4 is the absorbing state entered upon death; see Figure \ref{QoL}. 
The challenge in analysis of such data is to summarize the effect of treatment on the sojourn times in the different states or the state entry time distributions
in a robust way that facilitates simple comparisons between the treatment arms.
In \cite{Gelber1989}, and a series of papers that followed by Gelber and colleagues,  
this was achieved by assigning utilities to each day in each state and accumulating this over time. 
The summary measure was called the Quality-adjusted Time Without Symptoms of disease and Toxicity to treatment,  abbreviated as Q-TWiST \citep{Gelber1989}.  

In such a utility-based approach we specify a function $U(Z(t), t)$ which is the utility value or score (for example a cost or quality of life score) for an individual who is in state $Z(t)$ at time $t$.  The cumulative utility score up to time $\tau$ is  $U(\tau)= \int_{0}^{\tau} U(Z(t),t)dt$.  Often it is  assumed that $U(k, t) = U_{k}$  is a fixed 
utility rate for state $k$.  In this case the cumulative utility score  reduces to $U(\tau) = \sum_{k \in \mathcal{S}} \int_{0}^{\tau} U_{k} I(Z(t)=k)dt$, with expectation
\begin{equation}
\label{TotalUtilityScore}
\mu(\tau \ | \ X) =  \mathbb{E}(U(\tau)  \ | \ X) = \sum_{k \in \mathcal{S}} \int_{0}^{\tau} U_{k} P(Z(t)=k  \ | \ X)dt.
\end{equation} 
An estimand can be defined as $\beta(\tau) = \mu(\tau \ | \ X=1) - \mu(\tau \ | \ X=0)$ or alternatively as a ratio,  and can be given a descriptive causal interpretation 
because $U(\tau)$ is a marginal feature.  A straightforward approach is to use nonparametric Aalen-Johansen estimates of state occupancy probabilities $P(Z(t)=k|X)$ for each treatment group to give robust estimates for $\mu(\tau|X=1)-\mu(\tau|X=0)$ or for analogous ratios, with variance estimates  based on a nonparametric bootstrap. \cite{Cook2003} and \citet[Section 4.2]{Cook2018} illustrate this approach.
In spite of the challenge of specifying the utility functions \citep{Thomas1984, Torrance1986, Torrance1987}, 
this approach is attractive in balancing and combining multiple outcomes.  Utility-based secondary analysis can in addition be useful when considering the implications of treatment assignment.

{
We reiterate that the utility formulation $(\ref{TotalUtilityScore})$ covers features based on times to event $T_E$ as special cases where utility rates are either $0$ or $1$ for each state.  The restricted mean time to failure $RMT(\tau|X)$ is, for instance, given by (\ref{TotalUtilityScore}) with $U_0=1$, $U_1=U_2=0$.  The cumulative utility framework also includes the win ratio method as a special case.  In an illness-death process, for example, if $T_{1A}$ and $T_{2A}$ are the times of entry to states $1$ and $2$ for path A, and $T_{1B}$ and $T_{2B}$ are the respective entry times for path B,
then one rule would be to declare that path A is preferable (i.e. "wins") if and only if 
$T_{2B} < \min(T_{2A}, C)$ or 
$C < \min(T_{2A}, T_{2B})$ and $T_{1B}<\min(T_{1A},C)$, where $C$ is a common censoring time.  If we assign utility rates $U_0=1$ and $U_1= 1-\epsilon$ to states 0 and 1, with $\epsilon$ a small positive number, then the cumulative utility score up to time $C$ is $U(C)=W_0 + (1-\epsilon)W_1$, where $W_j$ is the total time spent in state $j \in \{0,  1\}$ over $[0, C]$. As $\epsilon$ approaches $0$ the probability that $U_A(C)>U_B(C)$ approximates the probability that A "wins". 
The utility-based approach is more flexible and tractable, however, and can handle problems such as variable and dependent censoring times $C$.
}

\section{Discussion} \label{sec6}
Randomized controlled trials that address disease processes are challenging to plan, conduct and analyze. In the simplest situations an intervention is designed to 
target a specific adverse event, but in many settings it might also affect other events. 
In this case there may be alternative choices for the primary event or outcome used for treatment comparisons. In addition, other disease-related events may prevent
 or interfere with the observation of the primary outcome,  or necessitate a change in treatment.  
A main message of this paper is the importance of formulating comprehensive models that incorporate features of the disease process as well as intercurrent events 
related to followup and interventions. 
Such models, in conjunction with medical information, aid in the formulation of protocols and specification of estimands.  
In addition, they are crucial for secondary  intensity-based analysis which provides a deeper understanding of treatment effects and interpretation of estimands 
based on marginal process features. 
{
The challenge of dealing with intercurrent events can sometimes be lessened by good study design and execution -- for example if one can follow individuals to an 
administrative censoring time then issues of loss to followup are avoided.
When the standard of care can be defined unambiguously problems in the interpretation of study findings might be avoided, but  in international multicenter trials
 investigators may view having variation in standard of care good for generalizability.
In such cases study findings are interpretable in a population with a similar mix of standard of care strategies, which may be difficult to use in a given setting.
}

Reference is often made to robustness and sensitivity analysis in conjunction with estimands and their estimation.  
Our view is that it is the model assumptions and probability limit $\beta^*$ of the estimator for an estimand $\beta$ that are crucial.  
For  causal interpretation of a marginal model-based  treatment effect  the model should be consistent with the observed data.  Models naturally just approximate reality and we prefer, 
as in some of our numerical illustrations,  to consider $\beta^*$ as the de facto estimand;  it can be interpreted  as the parameter value in the best 
approximating (or least false) model (see \citealp{Grambauer2010}).  
In this framework we would apply the term \textit{robust} to a family of marginal models and an estimand whose value $\beta^*$ does not change much under certain changes 
in the intensities for the disease or followup process.  
Numerical studies as we provided here can be viewed as sensitivity analyses concerning robustness.  
In this sense, the $cloglog$ transformation model for the cumulative incidence of a non-fatal event in competition with a fatal event is quite robust across a 
range of plausible scenarios, whereas the $log$ transformation model is less robust. 
Another aspect that is sometimes considered as a form of robustness, but  is slightly different, is whether a model is collapsible with respect to the 
effect of treatment when other covariates $V$ affecting treatment are removed (e.g. \citealp{Aalen2015}).  
This means that the treatment effect has the same interpretation in models with or without the inclusion of $V$.  
Aside from certain types of linear models, most models are not collapsible.  Proportional hazards and other relative risk models are in particular generally non-collapsible.

We note in the same vein that causal interpretation  of a marginal treatment effect does not rely on the "correctness" of the assumed model.  
It requires only that there is a well defined limiting estimand $\beta^*$ which is marginal in the sense that it is based on a marginal feature that involves no 
conditioning on events after randomization to treatment. 
An understanding of the treatment effect does however depend on model adequacy, as well as its source in the effects of treatment on process intensities. 

For disease processes associated with multiple types of events there may be no optimal primary outcome.  For example, in the LEADER trial the 
effects of treatment on cardiovascular events such as stroke, MI and CV death were of interest \citep{Furberg2021}.  
One approach to specification of a primary estimand is to consider a composite outcome, for example, the time to a first stroke, MI or CV death or the cumulative 
number of events of all types.  
Another is to use a cumulative utility score that recognizes the burden of various events or states.  
We encourage wider consideration of this approach, including discussion and research on utility scoring systems.  
Cumulative utilities can accommodate intercurrent events,  besides addressing  disease burden and quality of life.  This is facilitated by including states that deal with disease severity and effects of treatment in a model.  In an oncology trial involving alternative chemotherapies, we could for example split a relapse-free state into two states - one while a subject is undergoing chemotherapy and one for when toxic effects of the therapy have subsided \citep{Cook2003}. The first of the two states would have a lower utility (quality of life) score than the second. In a cardiovascular trial, the severity and after-effects of strokes or MIs could likewise be represented with multiple states, as opposed to simply recognizing the occurrence of the event.

Tests for treatment effects are a crucial aspect of randomized trials and inform decisions concerning sample sizes and trial duration. A thorough discussion of testing is beyond our present scope.  Our position is that tests should normally be related to estimands, and based on Wald or score statistics that employ robust variance estimates. This provides validity when the assumed model  is misspecified to some degree,  but in saying this we point out that the null hypothesis being tested is $H_0: \{ \beta^*=0 \}$.  A stronger null hypothesis is that treatment does not affect the disease process but it is usually unclear how it could be efficiently tested.  Moreover, in trials where intercurrent events or multiple disease-related events are common,  it would often be viewed as an unrealistic hypothesis.





\newpage

\appendix

\section{Limiting values of estimators under model misspecification \label{app1}}
\subsection{Impact of covariate omission in Cox intensity-based model}
\label{app-Cox}
To derive the limiting value $\phi_1^{*}$ given in Section \ref{sec2.3} we let $\lambda_{01}(t | X) = \tilde{\lambda}_{1}\exp(\phi_1 X)$ be the misspecified intensity model for the non-fatal event. The score equation under the Cox model is
$$ U(\phi_1) = \sum_{i=1}^{n} \int_{0}^{\tau}  \overline{Y}_{i}(t) \biggl \{ X_i - \dfrac{S^{(1)}(t, \phi_1)}{S^{(0)}(t, \phi_1)}  dN_{i1}(t) \biggr \}, $$ 
where $\overline{Y}_{i}(t) = Y_{i0}(t)I(\min(C, R_i) \geq t), dN_{i1}(t) = I(T_{i1} = t)$ and $S^{(r)}(t, \phi_1) = \sum_{i=1}^{n}  \overline{Y}_{i}(t) X_{i}^r \exp(\phi_1 X_i)$ for $r=0,1$.  The limiting value $\phi_1^{*}$ is the solution to the equation $\mathbb{E}(U(\phi_1))=0$, where  
$$\mathbb{E}(U(\phi_1)) =  \int_{0}^{\tau}  \biggl \{ \mathbb{E} \biggl( \sum_{i=1}^{n} \overline{Y}_{i}(t) X_i dN_{i1}(t) \biggr) - \dfrac{\mathbb{E}(S^{(1)}(t, \phi_1))}{\mathbb{E}(S^{(0)}(t, \phi_1))} \mathbb{E} \biggl( \sum_{i=1}^{n} \overline{Y}_{i}(t) dN_{i1}(t) \biggr) \biggr \}, $$
and where expectations are taken with respect to the true illness-death, censoring and covariate processes.  Since $\{Z(t), t>0 \}$ is independent of $R$ given $X$ and $V$,  $P(R > t |X, V)=P(R > t)$ and $X \perp V$ due to randomization,  the expectations needed to compute this are
\begin{align*} 
& \mathbb{E}(S^{(1)}(t, \phi_1)) = nP(R > t) P(X=1) \exp(\phi_1) \sum_{v \in \{0, 1\}} P(V=v) P(Z(t^-)=0 | X=1, V=v) \\ 
&\mathbb{E}(S^{(0)}(t, \phi_1)) = nP(R > t) \sum_{x \in \{0, 1\}} \sum_{v \in \{0, 1\}} \exp(\phi_1 x) P(X=x) P(V=v) P(Z(t^-)=0 | X=x, V=v)\\
&\mathbb{E} \biggl( \sum_{i=1}^{n} \overline{Y}_{i}(t) X_i dN_{i1}(t) \biggr) = nP(R > t)P(X=1) \cdot  \\
& \ \ \ \ \ \ \ \ \ \ \ \ \ \ \ \ \ \ \ \ \ \ \ \ \biggl[ \ \sum_{v \in \{0, 1\}} P(V=v) \lambda_{01}(t | X=1, V=v) P(Z(t^-)=0 | X=1, V=v) \ \biggl]  \\
&\mathbb{E} \biggl( \sum_{i=1}^{n} \overline{Y}_{i}(t) dN_{i1}(t) \biggr) = nP(R > t) \\
 & \ \ \ \ \ \ \ \ \ \ \ \ \ \ \ \ \ \ \ \ \ \ \ \ \biggl[ \ \sum_{x \in \{0, 1\}}\sum_{v \in \{0, 1\}} P(X=x) P(V=v) \lambda_{01}(t | X=x, V=v) P(Z(t^-)=0 | X=x, V=v)\ \biggl]. 
\end{align*} 

\subsection{Limiting value of the Fine-Gray estimator under misspecification}
\label{appendix-FG}
The inverse probability of censoring weighted estimating equations for $\beta$ under the Fine and Gray estimation method is 
\begin{align}
\label{score-FG}
U(\beta^{FG}) =  \sum_{i=1}^{n} \int_{0}^{\infty} w_{i}(t)Y_{i}^{\dagger}(t)\biggl[ X_{i} - \dfrac{S^{(1)}(t, \beta^{FG})}{S^{(0)}(t, \beta^{FG})} \biggl] dN_{i1}(t),
\end{align}
where $Y_{i}^{\dagger}(t)=I(T_{1i} \geq t),  dN_{1i}(t)=I(T_{1i}=t), w_i(t)= I(C_{i} \geq \min(T_{i}, t))/G_{i}(\min(T_{i}, t)), $ and $S^{(r)}(t, \beta^{FG}) = \sum_{i=1}^{n} w_{i}(t) Y_{i}^{\dagger}(t) X_{i}^{r}\exp(\beta X_{i})$ for $r=0,1$ \citep[Section 4.1]{Cook2018}. Here $C_i=\min(C, R_i)$ is the net censoring time for subject $i=1,2,...,n$ with $C$ the administrative censoring time and $R_i$ the random censoring time. Setting (\ref{score-FG}) equal to zero and solving gives $\widehat{\beta}^{FG}$ following estimation of the censoring distribution $G_i(t)=P(R_i > t)$.  The estimand $\beta^{FG*}$ is the limiting value solving 
\begin{align}
\label{limiting-value-FG}
\int_{0}^{\infty} \biggl\{ \mathbb{E}\biggl(\sum_{i=1}^{n} w_{i}(t) Y_{i}^{\dagger}(t) X_{i} dN_{1i}(t) \biggr) - \dfrac{s^{(1)}(t, \beta^{FG})}{s^{(0)}(t, \beta^{FG})} \mathbb{E}\biggl( \sum_{i=1}^{n} w_{i}(t) Y_{i}^{\dagger}(t)  dN_{1i}(t) \biggr) \biggr\} = 0,
\end{align}
where $s^{(r)}(t, \beta^{FG}) = \mathbb{E}(S^{(r)}(t, \beta^{FG}))$ for $r=0,1$ and where the expectations are taken with respect to the true competing risks, censoring and covariate processes. The stabilized Fine-Gray estimator $\widehat{\beta}^{FG}_{stab}$ and the corresponding estimand  $\beta^{FG*}_{stab}$ in Section \ref{sec3.2} can be found as the solutions to (\ref{score-FG}) and (\ref{limiting-value-FG}) with $w_{i}(t)$ replaced by $I(C_{i} \geq \min(T_{i}, t))G_{i}(t)/G_{i}(\min(T_{i}, t))$,  respectively \citep[Section 4]{Fine1999}. If the censoring model is correctly specified and thus $\widehat{G}_{i}(t) \longrightarrow G_{i}(t)$, the limiting value $\beta^{FG*}$ does not depend on the censoring distribution. The limiting value $\beta^{FG*}_{stab}$, on the other hand, varies with the censoring distribution due to weight stabilization. The Fine-Gray approach with weight stabilization is implemented in the R-function crr.

\subsection{Limiting value of the direct binomial estimator under misspecification}
\label{appendix-DB}
Using the same notation as in Section \ref{appendix-FG} and assuming estimation is conducted at several times $s_1 < s_2 < ... < s_R$ over the time horizon $[0, C]$,  the estimators $\widehat{\theta}=(\widehat{\alpha}_0(s_1),..., \widehat{\alpha}_0(s_R),\widehat{\beta}^{DB})$ are the solution to the generalized estimating equation \citep{Scheike2008}
\begin{align}
\label{limiting-value-DB}
U^{DB}(\theta) = \sum_{i=1}^{n} 
\begin{pmatrix}
    U_{i}(\alpha_{0}(s_1))  \\
    ...  \\
    U_{i}(\alpha_{0}(s_R))  \\
    U_{i}(\beta^{DB}) \\
\end{pmatrix}
= \sum_{i=1}^{n} D_{i} V^{-1}_{i} \Bigg[
\begin{pmatrix}
    \widetilde{N}_{1i}(s_1)  \\
    ...  \\
    \widetilde{N}_{1i}(s_R)  \\
\end{pmatrix} 
- 
\begin{pmatrix}
    F_{1}(s_1 | X_{i})  \\
    ...  \\
    F_{1}(s_R | X_{i}) \\
\end{pmatrix} \Bigg] = 
\begin{pmatrix}
    0 \\
    ...  \\
    0 \\
    0 \\
\end{pmatrix} \; , 
\end{align}
where $\theta=(\alpha_{0}(s_1), ..., \alpha_{0}(s_R), \beta^{DB})$ is a $(R+1)$-dimensional vector of unknown regression coefficients, $\widetilde{N}_{1i}(t) = I(C_{i} \geq \min(T_{i}, t))N_{1i}(t) / G_{i}(\min(T_i, t))$ is a weighted response, $F_{1}(s_r | X_{i}) = g^{-1}(\alpha_{0}(s_r) + \beta^{DB} X_{i})$ is the cumulative incidence function under the cloglog transformation model at time point $s_r$ ($r=1,...,R$),  $g(u)=\log(-\log(1-u))$ is the link function with $g^{-1}(u)=h(u)=1-\exp(-\exp(u))$, $V_{i}=\text{diag} \bigl( \{ F_1(s_r |X_{i})(1-F_{1}(s_r|X_{i})) \},  \ r=1,...,R\bigr)$ is a $(R \ \text{x} \ R)$ working independence covariance matrix, and $D_{i}$ is $((R+1) \ \text{x} \ R)$ matrix of derivatives of the form 
\begin{align*}
\begin{pmatrix}
    \dfrac{\partial F_{1}(s_1|X_i)}{\partial \alpha_{0}(s_1)} & 0  & ... &  ... &  0 &  0\\
    0 &  \dfrac{\partial F_{1}(s_2|X_i)}{\partial \alpha_{0}(s_2)} & 0 & ....  &  0 & 0\\
        ... & ... & ...  & ...  & ... &  .. . \\
    ... & ... & ...  & ...  & \dfrac{\partial F_{1}(s_{R-1}|X_i)}{\partial \alpha_{0}(s_{R-1})}  &  .. . \\
    0   & ... & ...  & ...  &   0  & \dfrac{\partial F_{1}(s_R|X_i)}{\partial \alpha_{0}(s_{R})}\\
    \dfrac{\partial F_{1}(s_1|X_i)}{\partial \beta^{DB}} & \dfrac{\partial F_{1}(s_2|X_i)}{\partial \beta^{DB}} & ...  & ... &  \dfrac{\partial F_{1}(s_{R-1}|X_i)}{\partial \beta^{DB}} &  \dfrac{\partial F_{1}(s_R|X_i)}{\partial \beta^{DB}}  \\
\end{pmatrix} \; , 
\end{align*}
with $\dfrac{\partial F_{1}(s_r|X_i)}{\partial \alpha_{0}(s_r)}=h'(\alpha_{0}(s_r)+\beta^{DB} X_{i})$ and $\dfrac{\partial F_{1}(s_r|X_i)}{\partial \beta^{DB}}=h'(\alpha_{0}(s_r)+\beta^{DB} X_{i})X_i$. This estimation procedure also requires modeling of the censoring distribution $G_i(t)=P(R_i > t)$. If the censoring model is correctly specified, the limiting values $\theta^{*} = (\alpha_0^{*}(s_1),..., \alpha_0^{*}(s_R),\beta^{DB*})$ can be found as the solution to $\mathbb{E}(U^{DB}(\theta))=(0, 0,..., 0)$, where 
\begin{align}
\label{score-DB1}
\mathbb{E}(U_{i}(\alpha_{0}(s_r)) &=  \sum_{x=0,1} \dfrac{P(X=x)h'(\alpha_{0}(s_r)+\beta^{DB}x)}{h(\alpha_0(s_r)+\beta^{DB} x)(1-h(\alpha_0(s_r)+\beta^{DB} x))} (F_1^{true}(s_r | X=x) - h(\alpha_0(s_r) + \beta^{DB} x)) \; , 
\end{align}
and
\begin{align}
\label{score-DB2}
\mathbb{E}(U_{i}(\beta^{DB})) &=  \sum_{s_r:  \ r=1,.., R} \dfrac{P(X=1)h'(\alpha_0(s_r)+\beta^{DB})}{h(\alpha_0(s_r)+\beta^{DB})(1-h(\alpha_0(s_r)+\beta^{DB}))} (F^{true}_1(s_r | X=1) - h(\alpha_0(s_r) + \beta^{DB}) ) \; . 
\end{align}
Here $F^{true}_1(s_r | X=1)$ is the true cumulative incidence function given in (\ref{CIF}) and the expectations are taken with respect to the true competing risks, censoring and covariate processes. 

\newpage 

\bibliographystyle{apalike}
\bibliography{main}  

\begin{thebibliography}{}

\bibitem[Aalen et~al., 2008]{Aalen2008}
Aalen, O., Borgan, {\O}., and Gjessing, H. (2008).
\newblock {\em {Survival and Event History Analysis: A Process Point of View}}.
\newblock Springer Science + Business Media, New York, NY.

\bibitem[Aalen, 2012]{Aalen2012}
Aalen, O.~O. (2012).
\newblock Armitage lecture 2010: Understanding treatment effects: the value of
  integrating longitudinal data and survival analysis.
\newblock {\em Statistics in Medicine}, 31(18):1903--1917.

\bibitem[Aalen et~al., 2015]{Aalen2015}
Aalen, O.~O., Cook, R.~J., and R{\o}ysland, K. (2015).
\newblock Does {C}ox analysis of a randomized survival study yield a causal
  treatment effect?
\newblock {\em Lifetime Data Analysis}, 21(4):579--593.

\bibitem[Aalen et~al., 2012]{aalen2012-2}
Aalen, O.~O., R{\o}ysland, K., Gran, J.~M., and Ledergerber, B. (2012).
\newblock Causality, mediation and time: a dynamic viewpoint.
\newblock {\em Journal of the Royal Statistical Society: Series A (Statistics
  in Society)}, 175(4):831--861.

\bibitem[Andersen et~al., 2019]{Andersen2019}
Andersen, P., Angst, J., and Ravn, H. (2019).
\newblock Modeling marginal features in studies of recurrent events in the
  presence of a terminal event.
\newblock {\em Lifetime Data Analysis}, 25:681--695.

\bibitem[Andersen et~al., 1993]{Andersen1993}
Andersen, P.~K., Borgan, O., Gill, R.~D., and Keiding, N. (1993).
\newblock {\em Statistical Models Based on Counting Processes}.
\newblock Springer-Verlag, New York, NY.

\bibitem[Andersen and Keiding, 2012]{Andersen2012.1}
Andersen, P.~K. and Keiding, N. (2012).
\newblock Interpretability and importance of functionals in competing risks and
  multistate models.
\newblock {\em Statistics in Medicine}, 31(11-12):1074--1088.

\bibitem[Barnett et~al., 1998]{Barnett1998}
Barnett, H., Taylor, D., Eliasziw, M., Fox, A., Ferguson, G., Haynes, R.,
  Rankin, R., Clagett, G., Hachiniski, V., Sackett, D., and Thorpe, K. (1998).
\newblock {Benefit of carotid endarterectomy in patients with symptomatic
  moderate or severe stenosis}.
\newblock {\em New England Journal of Medicine}, 339(20):1415--1425.

\bibitem[Booth and Eisenhauer, 2012]{booth2012}
Booth, C.~M. and Eisenhauer, E.~A. (2012).
\newblock Progression-free survival: meaningful or simply measurable?
\newblock {\em Journal of Clinical Oncology}, 30(10):1030--1033.

\bibitem[Bornkamp et~al., 2021]{bornkamp2021}
Bornkamp, B., Rufibach, K., Lin, J., Liu, Y., Mehrotra, D.~V., Roychoudhury,
  S., Schmidli, H., Shentu, Y., and Wolbers, M. (2021).
\newblock Principal stratum strategy: Potential role in drug development.
\newblock {\em Pharmaceutical Statistics}, 20(4):737--751.

\bibitem[B\"uhler et~al., 2022]{buhler22}
B\"uhler, A., Cook, R.~J., and Lawless, J.~L. (2022).
\newblock Semicompeting risks with recurrent events: marginal models and
  estimands.
\newblock {\em Manuscript}.

\bibitem[Carey et~al., 2021]{carey2021}
Carey, L., Loirat, D., Punie, K., Bardia, A., Dieras, V., Dalenc, F., Diamond,
  J., Fontaine, C., Wang, G., Rugo, H., Hurvitz, S., Kalinsky, K.,
  {O'Shaughnessy}, J., Loibl, S., Gianni, L., {Piccart-Gebhart}, M., Hong, Q.,
  Olivo, M., Itri, L., and Cortes, J. (2021).
\newblock {Assessment of sacituzumab govitecan (SG) in patients with prior
  neoadjuvant/adjuvant chemotherapy in the phase 3 ASCENT study in metastatic
  triple-negative breast cancer (mTNBC)}.
\newblock {\em Journal of Clinical Oncology}, 39(15 suppl):1080.

\bibitem[Casey et~al., 2021]{casey2021}
Casey, M., Degtyarev, E., Lechuga, M.~J., Aimone, P., Ravaud, A., Motzer,
  R.~J., Liu, F., Stalbovskaya, V., Tang, R., Butler, E., Sailer, O., Halabi,
  S., and George, D. (2021).
\newblock Estimand framework: Are we asking the right questions? a case study
  in the solid tumor setting.
\newblock {\em Pharmaceutical Statistics}, 20(2):324--334.

\bibitem[Coleman, 2006]{coleman2006}
Coleman, R. (2006).
\newblock {Clinical features of metastatic bone disease and risk of skeletal
  morbidity}.
\newblock {\em Clinical Cancer Research}, 12(20):6243s--6249s.

\bibitem[Comment et~al., 2019]{Comment2019}
Comment, L., Mealli, F., Haneuse, S., and Zigler, C. (2019).
\newblock Survivor average causal effects for continuous time: a principal
  stratification approach to causal inference with semicompeting risks.
\newblock https://doi.org/10.48550/arXiv.1902.09304.

\bibitem[Committee, 2017]{iche9-2017}
Committee, I. E.~S. (2017).
\newblock Ich e9 (r1): Addendum to statistical principles for clinical trials
  on choosing appropriate estimands and defining sensitivity analyses in
  clinical trials.
\newblock {\em International Conference on Harmonization}.

\bibitem[Cook et~al., 2009]{Cook2009}
Cook, R., Lawless, J., Lakhal-Chaieb, L., and Lee, K.-A. (2009).
\newblock {Robust estimation of mean functions and treatment effects for
  recurrent events under event-dependent censoring and termination: application
  to skeletal complications in cancer metastatic to bone}.
\newblock {\em Journal of the American Statistical Association},
  104(485):60--75.

\bibitem[Cook and Lawless, 2007]{Cook2007}
Cook, R.~J. and Lawless, J.~F. (2007).
\newblock {\em Statistical Analysis of Recurrent Events}.
\newblock Springer-Verlage, New York, NY.

\bibitem[Cook and Lawless, 2018]{Cook2018}
Cook, R.~J. and Lawless, J.~F. (2018).
\newblock {\em Multistate Models for the Analysis of Life History Data}.
\newblock CRC Press.

\bibitem[Cook et~al., 2003]{Cook2003}
Cook, R.~J., Lawless, J.~F., and Lee, K.-A. (2003).
\newblock Cumulative processes related to event histories.
\newblock {\em SORT}, 27(1):13--30.

\bibitem[Cook and Lawless, 2022]{Cook2022}
Cook, R.~J. and Lawless, J.~L. (2022).
\newblock Life history analysis: a review and some current issues.
\newblock {\em Canadian Journal of Statistics}.

\bibitem[Datta and Satten, 2002]{Datta2002}
Datta, S. and Satten, G. (2002).
\newblock {Estimation of integrated transition hazardsand stage occupation
  probabilities for non-Markov systems under dependent censoring}.
\newblock {\em Biometrics}, 58(4):792--802.

\bibitem[Fine and Gray, 1999]{Fine1999}
Fine, J.~P. and Gray, R.~J. (1999).
\newblock A {P}roportional {H}azards {M}odel for the {S}ubdistribution of a
  {C}ompeting {R}isk.
\newblock {\em Journal of the American Statistical Association},
  94(446):496--509.

\bibitem[Furberg et~al., 2022]{Furberg2021}
Furberg, J.~K., Rasmussen, S., Andersen, P.~K., and Ravn, H. (2022).
\newblock Methodological challenges in the analysis of recurrent events for
  randomised controlled trials with application to cardiovascular events in
  leader.
\newblock {\em Pharmaceutical Statistics}, 21(1):241--267.

\bibitem[Gelber et~al., 1995]{Gelber1995}
Gelber, R.~D., Cole, B.~F., Gelber, S., and Goldhirsch, A. (1995).
\newblock Comparing treatments using quality-adjusted survival: The q-twist
  method.
\newblock {\em The American Statistician}, 49(2):161--169.

\bibitem[Gelber et~al., 1989]{Gelber1989}
Gelber, R.~D., Gelman, R.~S., and Goldhirsch, A. (1989).
\newblock A quality-of-life-oriented endpoint for comparing therapies.
\newblock {\em Biometrics}, 45(3):781--795.

\bibitem[Gerds et~al., 2012]{Gerds2012}
Gerds, T.~A., Scheike, T.~H., and Andersen, P.~K. (2012).
\newblock Absolute risk regression for competing risks: interpretation, link
  functions, and prediction.
\newblock {\em Statistics in Medicine}, 31(29):3921--3930.

\bibitem[Ghosh and Lin, 2002]{Ghosh2002}
Ghosh, D. and Lin, D. (2002).
\newblock {Marginal regression models for recurrent and terminal events}.
\newblock {\em Statistica Sinica}, 12(3):663--688.

\bibitem[Glasziou et~al., 1990]{Glasziou1990}
Glasziou, P.~P., Simes, R.~J., and Gelber, R.~D. (1990).
\newblock Quality adjusted survival analysis.
\newblock {\em Statistics in Medicine}, 9:1259--1276.

\bibitem[Grambauer et~al., 2010]{Grambauer2010}
Grambauer, N., Schumacher, M., and Beyersmann, J. (2010).
\newblock Proportional subdistribution hazards modeling offers a summary
  analysis, even if misspecified.
\newblock {\em Statistics in Medicine}, 29(7--8):875--884.

\bibitem[Henshall et~al., 2016]{henshall2016}
Henshall, C., Latimer, N., Sansom, L., and Ward, R. (2016).
\newblock Treatment switching in cancer trials: issues and proposals.
\newblock {\em International Journal of Technology Assessment in Health Care},
  32(3):167--174.

\bibitem[Hern\'an and Robins, 2020]{hernan2020}
Hern\'an, M. and Robins, J. (2020).
\newblock {\em Causal Inference: What if}.
\newblock Boca Raton: Chapman \& Hill/CRC.

\bibitem[Hern\'an, 2010]{Hernan2010}
Hern\'an, M.~A. (2010).
\newblock The hazards of hazard ratios.
\newblock {\em Epidemiology}, 21(1):13--15.

\bibitem[Hern{\'a}n and Scharfstein, 2018]{hernan2018}
Hern{\'a}n, M.~A. and Scharfstein, D. (2018).
\newblock Cautions as regulators move to end exclusive reliance on intention to
  treat.
\newblock {\em Annals of Internal Medicine}, 168(7):515--516.

\bibitem[Jeong and Fine, 2007]{Jeong2007}
Jeong, J.-H. and Fine, J.~P. (2007).
\newblock Parametric regression on cumulative incidence function.
\newblock {\em Biostatistics}, 8(2):184--196.

\bibitem[Klein and Andersen, 2005]{Klein2005}
Klein, P.~J. and Andersen, P.~K. (2005).
\newblock Regression modeling of competing risks data based on pseudovalues of
  the cumulative incidence function.
\newblock {\em Biometrics}, 61(1):223--229.

\bibitem[Latimer et~al., 2019a]{latimer2019b}
Latimer, N., Abrams, K., and Siebert, U. (2019a).
\newblock Two-stage estimation to adjust for treatment switching in randomised
  trials: a simulation study investigating the use of inverse probability
  weighting instead of re-censoring.
\newblock {\em BMC Medical Research Methodology}, 19(1):1--19.

\bibitem[Latimer et~al., 2019b]{latimer2019}
Latimer, N., White, I., Abrams, K., and Siebert, U. (2019b).
\newblock Causal inference for long-term survival in randomised trials with
  treatment switching: Should re-censoring be applied when estimating
  counterfactual survival times?
\newblock {\em Statistical Methods in Medical Research}, 28(8):2475--2493.

\bibitem[Lawless and Cook, 2019]{Lawless2019}
Lawless, J. and Cook, R. (2019).
\newblock {A new perspective on loss to follow-up in failure time and life
  history studies}.
\newblock {\em Statistics in Medicine}, 38(23):4583--4610.

\bibitem[Lipkovich et~al., 2022]{lipkovich2022}
Lipkovich, I., Ratitch, B., Qu, Y., Zhang, X., Shan, M., and Mallinckrodt, C.
  (2022).
\newblock Using principal stratification in analysis of clinical trials.
\newblock {\em Statistics in Medicine}.

\bibitem[Llewellyn-Thomas et~al., 1984]{Thomas1984}
Llewellyn-Thomas, H.~A., Sutherland, H.~J., Tibshirani, R., Ciampi, A., Till,
  J.~E., and Boyd, N.~F. (1984).
\newblock Describing health states: Methodologic issues in obtaining values for
  health states.
\newblock {\em Medical Care}, 22:543--552.

\bibitem[{Ludwig Breast Cancer Study Group}, 1988]{LBCSG1988}
{Ludwig Breast Cancer Study Group} (1988).
\newblock {Combination adjuvant chemotherapy for node-positive breast cancer}.
\newblock {\em New England Journal of Medicine}, 319(11):677--683.

\bibitem[Luo et~al., 2015]{luo2015}
Luo, X., Tian, H., Mohanty, S., and Tsai, W.~Y. (2015).
\newblock An alternative approach to confidence interval estimation for the win
  ratio statistic.
\newblock {\em Biometrics}, 71(1):139--145.

\bibitem[Manitz et~al., 2022]{Manitz2022}
Manitz, J., Kan-Dobrosky, N., Buchner, H., Casadebaig, M.-L., Degtyarev, E.,
  Dey, J., Haddad, V., Jie, F., Martin, E., Mo, M., Rufibach, K., Shentu, Y.,
  Stalbovskaya, V., (Sammi)Tang, R., Yung, G., and Zhou, J. (2022).
\newblock Estimands for overall survival in clinical trials with treatment
  switching in oncology.
\newblock {\em Pharmaceutical Statistics}, 21(1):150--162.

\bibitem[Mao, 2019]{mao2019}
Mao, L. (2019).
\newblock On the alternative hypotheses for the win ratio.
\newblock {\em Biometrics}, 75(1):347--351.

\bibitem[Mao and Kim, 2021]{Mao2021b}
Mao, L. and Kim, K. (2021).
\newblock {Statistical models for composite endpoints of death and nonfatal
  events: a review}.
\newblock {\em Statistics in Biopharmaceutical Research}, 13(3):260--269.

\bibitem[Mao et~al., 2021]{mao2021}
Mao, L., Kim, K., and Miao, X. (2021).
\newblock Sample size formula for general win ratio analysis.
\newblock {\em Biometrics}.

\bibitem[Mao and Lin, 2016]{Mao2016}
Mao, L. and Lin, D. (2016).
\newblock {Semiparametric regression for the weighted composite endpoint of
  recurrent and terminal events}.
\newblock {\em Biostatistics}, 17(2):390--403.

\bibitem[Martinussen et~al., 2020]{Martinussen2020}
Martinussen, T., Vansteelandt, S., and Andersen, P.~K. (2020).
\newblock Subtleties in the interpretation of hazard contrasts.
\newblock {\em Lifetime Data Analysis}, 26(04):833--855.

\bibitem[Montori and Guyatt, 2001]{montori2001}
Montori, V. and Guyatt, G. (2001).
\newblock Intention-to-treat principle.
\newblock {\em CMAJ}, 165(10):1339--1341.

\bibitem[Nevo and Gorfine, 2021]{Nevo2020}
Nevo, D. and Gorfine, M. (2021).
\newblock {Causal inference for semi-competing risks data}.
\newblock {\em Biostatistics}.

\bibitem[{North America Symptomatic Carotid Endarterectomy Trial Steering
  Committee}, 1991]{NASCET-STROKE1991}
{North America Symptomatic Carotid Endarterectomy Trial Steering Committee}
  (1991).
\newblock {North America Symptomatic Carotid Endarterectomy Trial: methods,
  patient characteristics and progress}.
\newblock {\em Stroke}, 22:711--720.

\bibitem[{North American Symptomatic Carotid Endarterectomy Trial
  Collaborators}, 1991]{NASCET-NEJM1991}
{North American Symptomatic Carotid Endarterectomy Trial Collaborators} (1991).
\newblock {Beneficial effect of carotid endarterectomy in symptomatic patients
  with high-grade carotid stenosis}.
\newblock {\em New England Journal of Medicine}, 325(7):445--453.

\bibitem[Oakes, 2016]{oakes2016}
Oakes, D. (2016).
\newblock On the win-ratio statistic in clinical trials with multiple types of
  event.
\newblock {\em Biometrika}, 103(3):742--745.

\bibitem[Pocock et~al., 2012]{pocock2012}
Pocock, S.~J., Ariti, C.~A., Collier, T.~J., and Wang, D. (2012).
\newblock The win ratio: a new approach to the analysis of composite endpoints
  in clinical trials based on clinical priorities.
\newblock {\em European heart journal}, 33(2):176--182.

\bibitem[Powles et~al., 2018]{Powles2018}
Powles, T., Dur{\'a}n, I., {van der Heijden}, M., Loriot, Y., Vogelzang, N.,
  {De Giorgi}, U., Oudard, S., Retz, M., Castellano, D., Bamias, A.,
  Fl{\'e}chon, A., Gravis, G., Hussain, S., Takano, T., Leng, N., Kadel, E.,
  Banchereau, R., Hegde, P., Mariathasan, S., Cui, N., Shen, X., Derleth, C.,
  Green, M., and Ravaud, A. (2018).
\newblock Atezolizumab versus chemotherapy in patients with platinum-treated
  locally advanced or metastatic urothelial carcinoma (imvigor211): a
  multicentre, open-label, phase 3 randomised controlled trial.
\newblock {\em Lancet}, 391:748--757.

\bibitem[Poythress et~al., 2020]{Poythress2020}
Poythress, J.~C., Lee, M.~Y., and Young, J. (2020).
\newblock Planning and analyzing clinical trials with competing risks:
  Recommendations for choosing appropriate statistical methodology.
\newblock {\em Pharmaceutical Statistics}, 19(1):4--21.

\bibitem[Putter et~al., 2020]{Putter2020}
Putter, H., Schumacher, M., and van Houwelingen, H.~C. (2020).
\newblock On the relation between the cause-specific hazard and the
  subdistribution rate for competing risks data: The {F}ine--{G}ray model
  revisited.
\newblock {\em Biometrical Journal}, 62(3):790--807.

\bibitem[Qu et~al., 2021]{qu2021}
Qu, Y., Shurzinske, L., and Sethuraman, S. (2021).
\newblock Defining estimands using a mix of strategies to handle intercurrent
  events in clinical trials.
\newblock {\em Pharmaceutical Statistics}, 20(2):314--323.

\bibitem[Rittmeyer et~al., 2017]{Rittmeyer2017}
Rittmeyer, A., Barlesi, F., Waterkamp, D., Park, K., Ciardiello, F., {von
  Pawel}, J., Gadgeel, S., T., H., Kowalski, D., Dols, M., Cortinovis, D.,
  Leach, J., Polikoff, J., Barrios, C., Kabbinavar, F., Frontera, O., {De
  Marinis}, F., Turna, H., Lee, J., Ballinger, M., Kowanetz, M., He, P., Chen,
  D., Sandler, A., Gandara, D., and {for the OAK Study Group} (2017).
\newblock {Atezolizumab versus docetaxel in patients with previously treated
  non-small-cell lung cancer (OAK): a phase 3, open-label, multicentre
  randomised controlled trial}.
\newblock {\em The Lancet}, 389:255--265.

\bibitem[Rubin, 2005]{rubin2005}
Rubin, D.~B. (2005).
\newblock Causal inference using potential outcomes: Design, modeling,
  decisions.
\newblock {\em Journal of the American Statistical Association},
  100(469):322--331.

\bibitem[Rubin, 2006]{rubin2006}
Rubin, D.~B. (2006).
\newblock Causal inference through potential outcomes and principal
  stratification: application to studies with "censoring" due to death.
\newblock {\em Statistical Science}, pages 299--309.

\bibitem[Rufibach, 2019]{Rufibach2019}
Rufibach, K. (2019).
\newblock Treatment effect quantification for time-to-event endpoints --
  estimands, analysis strategies, and beyond.
\newblock {\em Pharmaceutical Statistics}, 18(2):145--165.

\bibitem[Scharfstein, 2019]{scharfstein2019}
Scharfstein, D.~O. (2019).
\newblock A constructive critique of the draft ich e9 addendum.
\newblock {\em Clinical Trials}, 16(4):375--380.

\bibitem[Scheike and Zhang, 2007]{Scheike2007}
Scheike, T. and Zhang, M.-J. (2007).
\newblock Direct modelling of regression effects for transition probabilities
  in multistate models.
\newblock {\em Scandinavian Journal of Statistics}, 34(1):17--32.

\bibitem[Scheike et~al., 2008]{Scheike2008}
Scheike, T.~H., Zhang, M.-J., and Gerds, T.~A. (2008).
\newblock Predicting cumulative incidence probability by direct binomial
  regression.
\newblock {\em Biometrika}, 95(1):205--220.

\bibitem[Schmidli et~al., 2021]{Schmidli2021}
Schmidli, H., Roger, J.~H., and Akacha, M. (2021).
\newblock Estimands for recurrent event endpoints in the presence of a terminal
  event.
\newblock {\em Statistics in Biopharmaceutical Research}, 0(0):1--11.

\bibitem[Snowden et~al., 2020]{snowden2020}
Snowden, J.~M., Reavis, K.~M., and Odden, M.~C. (2020).
\newblock Conceiving of questions before delivering analyses: relevant question
  formulation in reproductive and perinatal epidemiology.
\newblock {\em Epidemiology}, 31(5):644--648.

\bibitem[Stensrud and Dukes, 2022]{stensrud2021}
Stensrud, M.~J. and Dukes, O. (2022).
\newblock Translating questions to estimands in randomized clinical trials with
  intercurrent events.
\newblock {\em Statistics in Medicine}, 41(16):3211--3228.

\bibitem[Stensrud et~al., 2020]{Stensrud2020}
Stensrud, M.~J., Young, J.~G., Didelez, V., Robins, J.~M., and Hern\'an, M.~A.
  (2020).
\newblock Separable effects for causal inference in the presence of competing
  events.
\newblock {\em Journal of the American Statistical Association}, pages 1--23.

\bibitem[Ster et~al., 2020]{ster2020}
Ster, A. M.~C., Cornelius, V., and Cro, S. (2020).
\newblock Current approaches to handling rescue medication in asthma and eczema
  randomized controlled trials are inadequate: a systematic review.
\newblock {\em Journal of Clinical Epidemiology}, 125:148--157.

\bibitem[Struthers and Kalbfleisch, 1986]{Struthers1986}
Struthers, C.~A. and Kalbfleisch, J.~D. (1986).
\newblock Misspecified proportional hazard models.
\newblock {\em Biometrika}, 73(2):363--369.

\bibitem[Theriault et~al., 1999]{Theriault1999}
Theriault, R., Lipton, A., Hortobagyi, G., Leff, R., Gl{\"u}ckS., Stewart, J.,
  Costello, S., Kennedy, I., Simeone, J., Seaman, J., and Knight, R. (1999).
\newblock {Pamidronate reduces skeletal morbidity in women with advanced breast
  cancer and lytic bone lesions: a randomized, placebo-controlled trial}.
\newblock {\em Journal of Clinical Oncology}, 17(3):846--846.

\bibitem[Toenges et~al., 2021]{Toenges2021}
Toenges, G., M{\"{u}}tze, T., and {Jahn-Eimermacher}, A. (2021).
\newblock {A comparison of semiparametric approaches to evaluate composite
  endpoints in heart failure trials}.
\newblock {\em Statistics in Medicine}, 40(26):5702--5724.

\bibitem[Torrance, 1986]{Torrance1986}
Torrance, G.~W. (1986).
\newblock Measurement of health state utilities for economic appraisal: A
  review.
\newblock {\em Journal of Health Economics}, 5(1):1--30.

\bibitem[Torrance, 1987]{Torrance1987}
Torrance, G.~W. (1987).
\newblock Utility approach to measuring health-related quality of life.
\newblock {\em Journal of Chronic Diseases}, 40(6):593--600.

\bibitem[Watkins et~al., 2013]{watkins2013}
Watkins, C., Huang, X., Latimer, N., Tang, Y., and Wright, E. (2013).
\newblock Adjusting overall survival for treatment switches: commonly used
  methods and practical application.
\newblock {\em Pharmaceutical Statistics}, 12(6):348--357.

\bibitem[Wei et~al., 2021]{wei2021}
Wei, J., M{\"{u}}tze, T., {Jahn-Eimermacher}, A., and Roger, J. (2021).
\newblock {Properties of two while-alive estimands for recurrent events and
  their potential estimators}.
\newblock {\em Statistics in Biopharmaceutical Research}, pages 1--11.

\bibitem[Xu et~al., 2020]{Xu2020}
Xu, Y., Scharfstein, D., M\"uller, P., and Daniels, M. (2020).
\newblock A bayesian nonparametric approach for evaluating the causal effect of
  treatment in randomized trials with semi-competing risks.
\newblock {\em Biostatistics}.

\bibitem[Young et~al., 2020]{Young2020}
Young, J.~G., Stensrud, M.~J., Tchetgen~Tchetgen, E.~J., and Hern\'an, M.~A.
  (2020).
\newblock A causal framework for classical statistical estimands in
  failure-time settings with competing events.
\newblock {\em Statistics in Medicine}, 39(8):1199--1236.

\bibitem[Zhang and Fine, 2008]{Zhang2008}
Zhang, M.-J. and Fine, J. (2008).
\newblock Summarizing differences in cumulative incidence functions.
\newblock {\em Statistics in Medicine}, 27(24):4939--4949.

\end{thebibliography}

\end{document}